\documentclass[pdflatex,sn-nature]{sn-jnl}% Style for submissions to Nature Portfolio journals
\usepackage{graphicx}%
\usepackage{multirow}%
\usepackage{amsmath,amssymb,amsfonts}%
\usepackage{amsthm}%
\usepackage{mathrsfs}%
\usepackage[title]{appendix}%
\usepackage{xcolor}%
\usepackage{textcomp}%
\usepackage{manyfoot}%
\usepackage{booktabs}%
\usepackage{algorithm}%
\usepackage{algorithmicx}%
\usepackage{algpseudocode}%
\usepackage{listings}%
\usepackage{hyperref}
\usepackage{subfig,float}
\usepackage{cleveref}
\theoremstyle{thmstyleone}%
\theoremstyle{thmstyletwo}%

\theoremstyle{thmstylethree}%

\begin{document}

\title[]{Quantum annealers as programmable thermal machines}

\author[1,2]{Jakub Pawłowski}
\author[3]{Tomasz Śmierzchalski}
\author[4]{Fengping Jin}
\author[3]{Bartłomiej Gardas}
\author[5,6,7]{Sebastian Deffner}
%\author[4,8]{Kristel Michielsen}
\author*[4,8]{\fnm{} \sur{Zakaria Mzaouali}}\email{z.mzaouali@extern.fz-juelich.de}

\affil[1]{Institute of Theoretical Physics, Faculty of Fundamental Problems of Technology, Wrocław University of Science and Technology, 50-370, Wrocław, Poland}
\affil[2]{Quantumz.io Sp. z o.o., Puławska 12/3, 02-566, Warsaw, Poland}
\affil[3]{Institute of Theoretical and Applied Informatics, Polish Academy of Sciences, Bałtycka 5, Gliwice, 44-100, Poland}
\affil[4]{J\"ulich Supercomputing Centre, Institute for Advanced Simulation, Forschungszentrum J\"ulich, Wilhelm-Johnen-Straße, J\"ulich, 52428, Germany}
\affil[5]{Department of Physics, University of Maryland, Baltimore County, Baltimore, MD 21250, USA}
\affil[6]{National Quantum Laboratory, College Park, MD 20740, USA}
\affil[7]{Quantum Science Institute, University of Maryland, Baltimore County, Baltimore, MD 21250, USA}
\affil[8]{Institut für Theoretische Physik, Universität Tübingen, Auf der Morgenstelle 14, 72076 Tübingen, Germany}
%%==================================%%

% \abstract{
% Programmable quantum annealers are primarily evaluated by their solution quality, sampling capabilities, and time-to-solution in complex optimization tasks. However, these architectures fundamentally represent highly controlled, dissipative open quantum systems, offering a platform for investigating finite time quantum thermodynamics. Here, we demonstrate the realization of closed thermodynamic cycles using reverse annealing on D-Wave quantum processors. By measuring the two-point energy statistics of programmed Ising Hamiltonians—without the need for direct environmental access—we infer effective environment temperatures and establish bounds on heat, work, entropy production, and power. Systematically tuning the initial Gibbs ensemble and the reverse annealing turning point allows us to experimentally map the operational phase space of thermal machines, distinguishing heater-, accelerator-, refrigerator-, and engine-compatible regimes across one and two dimensional instances. Furthermore, we show that repeated execution of these cycles on a calibrated instance transforms the processor into an effective thermometer for the sampled computational degrees of freedom. Our results bridge the gap between quantum computation and non-equilibrium thermodynamics, establishing present day quantum annealers as versatile testbeds for thermal machine simulation, in situ thermometry, and energy statistics.}

\abstract{
Programmable quantum annealers are used for optimization, probabilistic
sampling, and simulation, but their performance is commonly reported without
the energy exchanged during computation. Here we characterize the D-Wave quantum annealer as a closed thermodynamic cycle. From initial
and final Ising energies and an effective temperature fitted to the output
distribution, we obtain lower bounds on entropy production, environment
energy exchange, work, and power. By
varying the prepared distribution and the reverse annealing turning point, we
map heater-, accelerator-, refrigerator-, and engine-compatible regimes in
one dimensional chains and higher connectivity instances, and apply the same
analysis to Advantage and Advantage2 hardware. For an encoded optimization
problem, the measured processor energy change states whether final candidates
improve or worsen the programmed objective on average. For sampling, the
fitted temperature provides an operational measure of how strongly
probability is concentrated among low energy configurations. The
thermodynamic mode therefore adds information absent from solution quality or
runtime alone: it distinguishes driven refinement, net heating, and heat
pumping while quantifying their energetic consequences. This framework
connects quantum optimization, probabilistic computing, statistical physics
simulation, hardware diagnostics, and energy-aware assessment without
assuming that a thermodynamic label alone determines computational
performance.
}


\maketitle

\section{Introduction}
Quantum annealers are routinely benchmarked by solution quality, sampling capabilities, and time-to-solution~\cite{HanussekPRAp2026, jakub_PRAp2026, TuzPRAp2026, Bando2020Universality, Sathe2026ClassicalCriticality, King2025BeyondClassical}. Much less is known about how their programmable schedules shape energy exchange, dissipation, and effective heat flow at the chip level~\cite{gardas2018quantum, Nelson_2022, Amin_2015, Marshall_2019, Kadowaki_2019}. Reverse annealing experiments on D-Wave processors have already shown that the device operates as a bona fide quantum thermal machine: during a cycle the chip absorbs and releases heat while external drives do work, and its net operation can be classified thermodynamically. In particular, previous work established that reverse annealing realizes a thermal accelerator cycle and provided quantitative, experimentally accessible bounds on heat, work, and dissipation using exact non-equilibrium relations~\cite{buffoni2020, Campisi_2021, Smierzchalski_2024, DoucetNJP}. These studies placed the D-Wave platform squarely within quantum thermodynamics and demonstrated that energy flows on the chip can be measured and constrained without full access to the environment~\cite{Deffner2019, blok2025quantum, Fabrizio2024}. 

Characterizing a D-Wave processor as a programmable thermal machine is relevant
because reverse annealing implements controlled, closed cycles in a driven open
many-spin system. This viewpoint extends the assessment of quantum annealers beyond
solution quality and time to solution by examining how the initial ensemble and the
annealing schedule determine energy exchange, entropy production, and work supplied
during the cycle. It therefore allows heater, accelerator, refrigerator, and
engine-compatible operation to be identified from energy statistics available at the
user level, while providing an experimental setting for testing finite time
thermodynamic relations in systems larger than typical few-qubit demonstrations. The
same framework also enables annealing protocols to be compared according to their
thermodynamic bounds as well as their computational output. In addition, fitting the
output configurations to the programmed Ising Hamiltonian yields an instance- and
protocol-dependent effective temperature of the sampled degrees of freedom, providing
a built-in thermometry diagnostic that complements, but does not replace, the
temperature measured by sensors in the dilution refrigerator. Treating the annealer as
a thermal machine therefore connects control, sampling, and energy exchange, and
provides a practical basis for characterizing and comparing quantum annealing
protocols and hardware
~\cite{Benedetti2016EffectiveTemperature,Raymond2016GlobalWarming,
Nelson2022HighQualityGibbs,Grattan2025ClassicalThermometry}.
\begin{figure*}
    \centering
    \includegraphics[width=\linewidth]{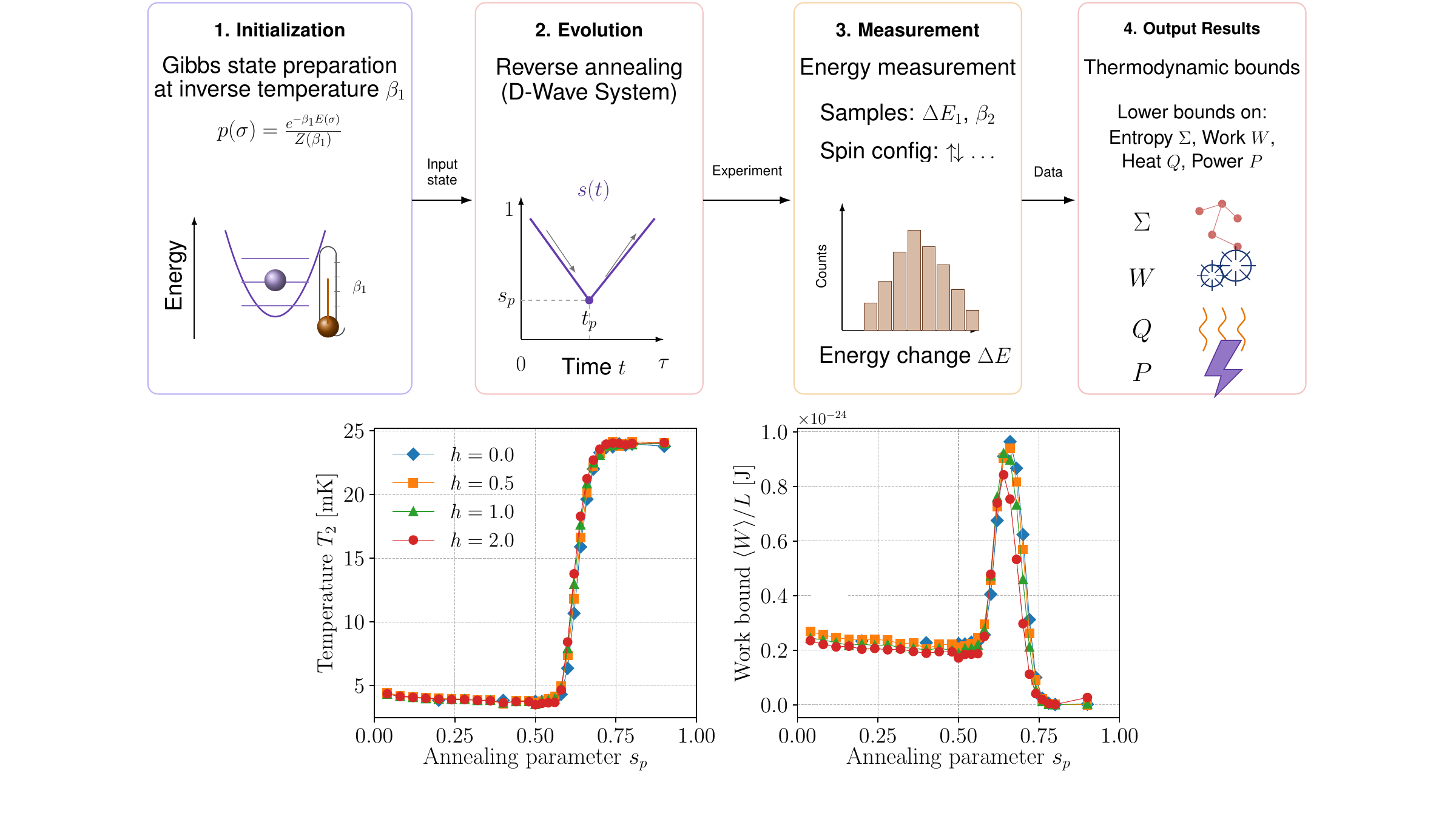}
    \caption{\textbf{Experimental workflow of the paper.}
A classical Gibbs state of the programmed Ising Hamiltonian is prepared at unitless inverse temperature $\beta_1=1$ and used as the input state for a reverse annealing schedule. The annealing parameter is ramped from $s=1$ to a turning point $s_{\rm p}$ and back to $s=1$, so that the initial and final Hamiltonians are the same. The quantum processing unit readout therefore gives a two point energy change $\Delta E_1$ of the problem Hamiltonian. Repeating the cycle builds the distribution of $\Delta E_1$, from which we estimate an effective environment temperature $T_2$ by pseudo-likelihood thermometry and evaluate fluctuation relation and thermodynamic uncertainty relation bounds on entropy production $\Sigma$, heat $Q$, work $W$ and power $P$. The procedure uses only experimentally accessible spin configurations and the programmed couplings and fields. The lower plots shows the estimated environment temperature and the lower bound on the work per spin for an Ising chain of length $N{=}300$, with random couplings drawn uniformly from $[-1,1]$ and fields from $[-h,h]$. The experiment was performed on the D-Wave Advantage5.4 system.}
\label{methods}
\end{figure*}
At the same time, there is growing interest in energy benchmarking of quantum hardware~\cite{Alexia2022,campbell2025roadmap, lipka2024thermodynamic, grattan2025classicalthermometryquantumannealers}. The efficiency of a quantum computer is ultimately limited by how it processes energy—both at cryogenic infrastructure level and at the chip level—so methods that tie computational performance to thermodynamic cost are essential~\cite{fellous2023, Linpeng2024, dassonneville2025amplifying, Linpeng2022, Stevens2022, maffei2023energy}. Recent experiments have begun to connect success probabilities and schedule design (e.g., reverse annealing with pauses~\cite{Pausing2020, Pausing2022, Vrinda2025, Vrinda2025bis}) to energetic figures of merit, again using thermodynamic bounds derived from fluctuations to turn limited measurements (energy changes reported by the quantum processing unit) into information about heat, work, and entropy production~\cite{Ishida2025}. This line of work argues for a unified view in which algorithmic choices and hardware schedules are co-designed for both accuracy and energy efficiency~\cite{Alexia2022, Meier2025}. 

Here we extend this thermodynamic perspective in two directions. First, we show that thermodynamic uncertainty relations (TURs) allow an extended thermodynamic description of the D-Wave quantum annealer beyond the accelerator thermal machine~\cite{buffoni2020}. In fact, TURs also allow the D-Wave quantum annealer to be described by the engine, heater, and refrigerator operation—by appropriate choices of schedule and initialization. We identify each regime directly from the signs of average energy changes of the processor and its environment over a cycle, and we implement protocols that realize all four behaviours on the same device. This demonstrates that the chip’s thermodynamic role is programmable: by steering the annealing path and the initial state, one can smoothly switch between accelerating heat flow, extracting useful work (engine), and pumping heat against a gradient (refrigerator). In doing so we complete, on real hardware, the thermodynamic classification previously proposed for annealing cycles~\cite{Campisi_2021}. Second, we leverage TURs to place a power bound on the D-Wave quantum annealer. Because only the processor’s energy change is directly observed, exact fluctuation identities together with TURs translate fluctuations in those changes into rigorous lower bounds on entropy production and into bounds on the average heat and work exchanged per cycle. Our approach yields practical, device agnostic limits on how much useful output any annealing schedule can deliver per unit time, and thus a quantitative route to discussing the energy efficiency of quantum annealers. 

Figure~\eqref{methods} summarizes the workflow of the paper used to probe the energy footprint of the D‑Wave quantum annealers by treating it as a thermal machine. By running reverse annealing cycles—where the system is first prepared in a classical state, partially annealed, then returned—we obtained data that tie the device’s dynamics directly to thermodynamics. We estimate four key quantities, using TURs as detailed in the Methods section~\eqref{methods}, for each problem instance: work done on the qubits, heat exchanged, entropy produced inside the chip, and the effective bath temperature that drives the process. We carried out the experiments on both one dimensional chains and two dimensional lattice problems that span a representative range of sizes and coupling strengths. Beyond their immediate implications for annealing, our results contribute to a broader goal of leveraging quantum thermodynamics—with fluctuation theorems and uncertainty relations as core tools—to assess and optimize emerging quantum technologies~\cite{Goold_2016, landi2021, koslov2014, Alicki1979TheQO, BENENTI20171, NMMAayers, Cangemi_2024}. By demonstrating that a commercial quantum annealer can be characterized by different thermal machine archetypes and by certifying performance limits from fluctuations alone, we provide a general blueprint for thermodynamics-inspired control of quantum hardware. This helps bridge foundational advances in non-equilibrium quantum physics with practical metrics for scalable, energy efficient quantum computation. 
\section{Results}
\paragraph*{Thermodynamics of quantum annealing.}
We programmed a nearest neighbour Ising chain of length $N{=}300$, with random couplings drawn uniformly from $[-1,1]$ and fields from $[-h,h]$, and started the processor in a Gibbs state of $H_p$ at inverse temperature $\beta_1{=}1$. The lower plots of figure~\eqref{methods} summarizes the environment temperature estimated from the pseudo-likelihood, Eq.~\eqref{beta2_est}, together with our TUR-based bounds on per-spin work, Eq.~\eqref{eq:TURW}, as functions of the annealing parameter $s_p$, for different longitudinal fields $h$, and at annealing time of $\tau = 100~\mu s$. The reported bath temperature $T_2$ lies in the $5$–$25$\,mK range across the anneal and exhibits a clear feature around $s_p\!\approx\!0.6$ where it rises and then levels off; the same $s_p$-window coincides with the peak in the (lower) bounds on work per spin and on power per spin.

The temperature extracted in the lower panel of Figure~\eqref{methods} should be interpreted as an effective temperature of the sampled Ising degrees of freedom, not as a direct measurement of
the refrigerator plate temperature. This distinction is important because the
pseudo likelihood, Eq.~\eqref{beta2_est}, results from the physical temperature, the programmed
energy scale, freeze-out effects, calibration errors, and residual non-equilibrium
dynamics. Its value is useful for the TUR-based characterization: when
the same instance and schedule are repeated, changes in \(T_2\) provide a sensitive
diagnostic of where the processor exchanges energy most strongly with its
environment.

This observation suggests a simple built-in thermometry protocol. One may choose a
calibrated reference instance, prepare the same initial Gibbs ensemble, repeat the
reverse annealing cycle, and infer \(T_2\) from the output bitstrings using the known
programmed Hamiltonian. Drifts in the inferred temperature then indicate changes in
the effective thermal state sampled by the active qubits. This does not replace
cryogenic thermometry, which measures the temperature of the surrounding hardware,
but it gives a local thermometer for the computational degrees of freedom that enter
the annealing experiment.

D-Wave documentation states that the quantum processing unit generally operates below $20$\,mK; our $T_2$ estimates fall within that envelope (allowing for calibration and the fact that our estimator is an effective temperature derived from qubit statistics rather than a direct thermometer). Additionally, for a one dimensional transverse field Ising model the quantum critical point occurs at $\Gamma/J{=}1$~\cite{pfeuty1970one}. When the anneal is written as $H(s)=A(s)H_x+B(s)H_p$ with $J$ normalized to unity, the critical region is reached when $A(s)/B(s)\!\approx\!1$, which on D-Wave systems occurs near the mid-anneal “crossing” of $A$ and $B$. In this region one expects enhanced susceptibility and energy fluctuations. Correspondingly, small changes in $s$ produce large changes in relaxation rates, effective temperature estimates, and dissipation. The sharp change we see around $s_p\!\approx\!0.6$ in $T_2$, together with the concomitant peaks in the lower bounds on work, is expected to match this critical point phenomenology. It also aligns with prior D-Wave experiments showing that dynamics and thermal repopulation are most pronounced near and just after the minimum gap, where pauses most effectively alter outcomes~\cite{Smierzchalski_2024}. 
\begin{figure*}
    \centering
    \includegraphics[width=\linewidth]{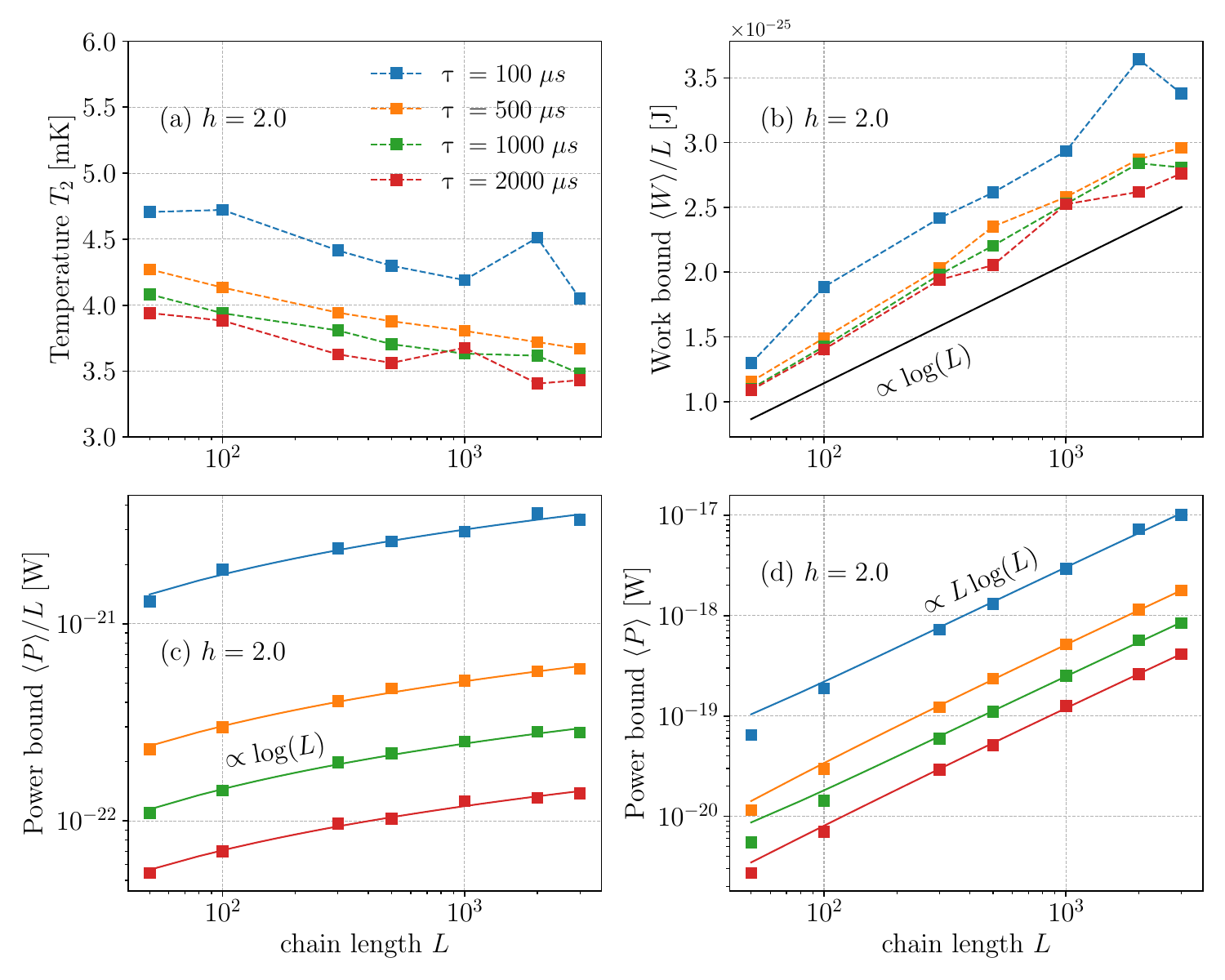}
    \caption{\textbf{Finite size scaling of thermodynamic bounds in a one dimensional chain.}
For random uniform chains with strong random fields, $h_i\in[-2,2]$, the reverse annealing turning point is fixed at $s_{p}=0.35$ and the chain length $L$ is varied on the D-Wave Advantage6.4 (Pegasus) chip. Panel (a) shows the pseudo-likelihood estimate of the effective temperature $T_2$ for annealing times $\tau=100$, $500$, $1000$ and $2000\,\mu\mathrm{s}$. Panels (b) and (c) show the work and power bounds per spin, which grow slowly with $L$ and are consistent with a logarithmic finite size enhancement over the range studied. Panel (d) shows the total power bound, which grows approximately as $L\log L$ when the per-spin contribution is summed over the chain.}
\label{fig_1D_random}
\end{figure*}
Quantitatively, the work bound per annealing cycle time gives an estimate on the power as defined in Eq.~\eqref{eq:power} of the Methods section~\eqref{methods}. The bound on the power follows the same trend as the work and peaks at $\mathcal{O}(10^{-21})$\,Watt, so even at the maximum we infer a power of order $10^{-19}$\,W for $N{=}300$---several orders of magnitude below the cryogenic overhead and fully compatible with sub-$20$\,mK operation. While this is a \emph{lower} bound on power (the true power can only be larger), it still provides a meaningful floor on dissipation that is physically small. D-Wave positions the annealing hardware as operating at very low temperatures with increased energy scale and reduced noise in newer generations; our observation that certified dissipation remains tiny—yet becomes measurably larger in the critical window where the system is most dynamically active—is consistent with those claims and with the expected concentration of energy exchange near the crossing of $A(s)$ and $B(s)$.

Our data support the following picture: (i) across the annealing parameter $s$ the environment behaves as an effectively cold bath in the millikelvin regime consistent with the chip specifications; (ii) near $s_p\!\approx\!0.6$ the chain is expected to traverses its critical region (defined by $A(s)/B(s)\!\approx\!1$), which amplifies energy fluctuations and irreversibility; and (iii) this amplification manifests simultaneously in $T_2$ and in the TUR bounds on work and power. The co-occurrence of these signatures at the same $s_p$ is the expected thermodynamic hallmark of the quantum critical point for this instance. 

\paragraph{Scaling with system size.}
For the one dimensional random uniform Ising instance (open chain with random couplings \(J_{ij}\in[-1,1]\), longitudinal field strength \(h_i\in[-2,2]\), initial Gibbs preparation at $\beta_1{=}1$) we observe that the TUR bounds on total work, Eq.~\eqref{eq:TURW}, and on power, Eq.~\eqref{eq:power}, grow as $L\log L$, while the corresponding \emph{per–spin} quantities grow as $\log L$ as shown in Figure~\eqref{fig_1D_random}. This scaling has a clear physical origin~\cite{pfeuty1970one, Campostrini2014, Sachdev_2011, Liu_2024}. The reverse annealing cycle traverses the critical region of the transverse field Ising model, where low momentum modes become dense and long ranged correlations develop. In that regime the relevant response functions are dominated by the infrared part of the spectrum: with a linear low–$k$ dispersion $\omega_k\simeq v|k|$ and an energy density correlator that yields an integrand $\propto 1/\omega_k$ (or, equivalently, a susceptibility kernel $\propto 1/|k|$), the finite size integral over available modes produces a logarithmic enhancement,
\begin{equation}
    \int_{k_{\min}}^{\Lambda}\frac{dk}{|k|}\; \sim\; \log\!\left(\frac{\Lambda}{k_{\min}}\right) \;\sim\; \log L,
\end{equation}
since $k_{\min}\!\sim\!\pi/L$. In linear response language, the excess (dissipative) work accumulated along the protocol is proportional to the time integral of such response kernels, so the per–spin dissipated work inherits the $\log L$ factor; multiplying by $L$ then gives the observed $L\log L$ scaling for the total bound. The same reasoning applies to the bound on power because the cycle time is held fixed across sizes, making power proportional to the bounded work.
\begin{figure*}
    \centering
    \includegraphics[width=\linewidth]{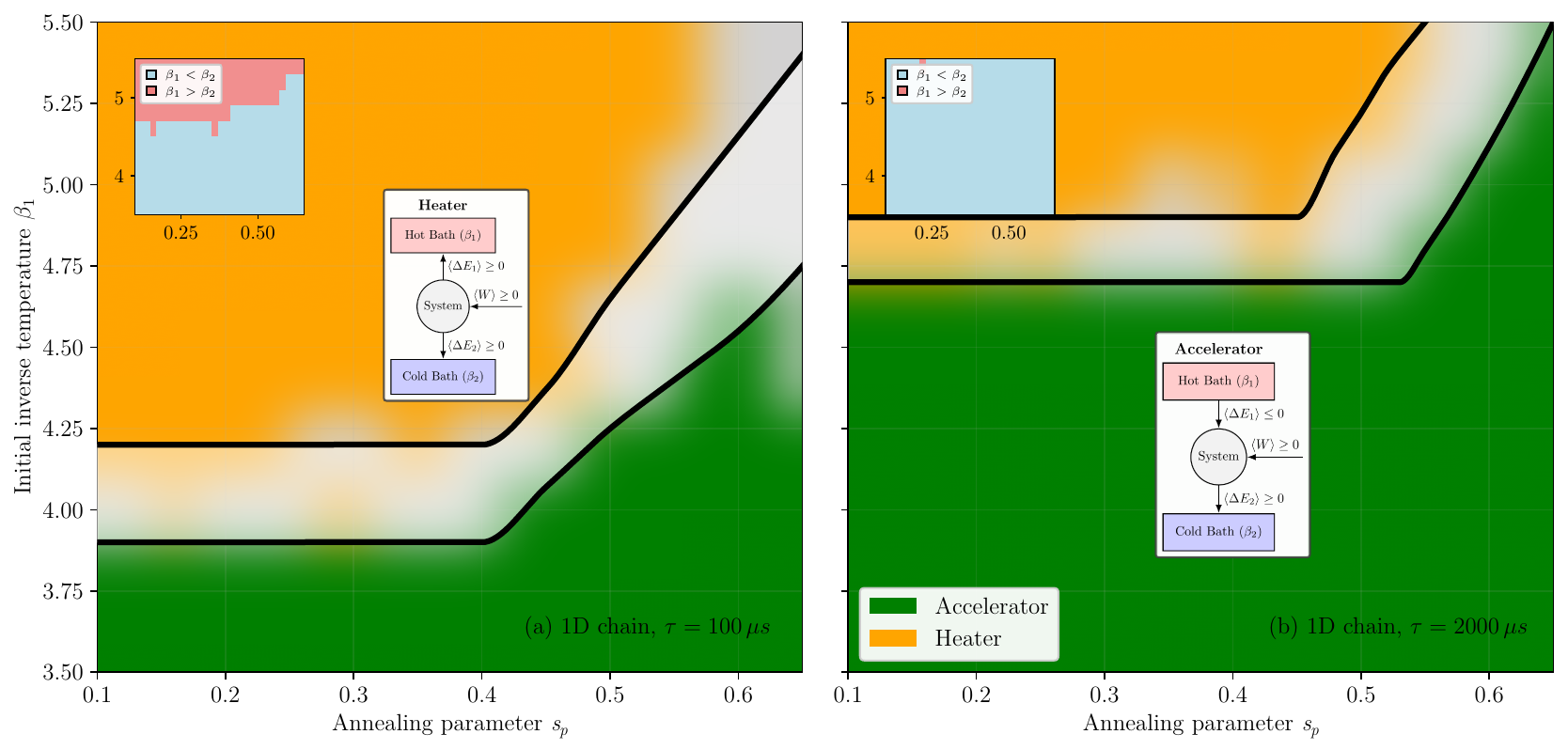}
    \caption{\textbf{Thermodynamic phase diagram of a one dimensional random uniform instance.}
Operating modes are mapped in the plane spanned by the initial inverse temperature $\beta_1$ and the reverse annealing turning point $s_{\rm p}$ for a one dimensional random uniform Ising instance. The two panels compare short and long cycles: (a) $\tau=100\,\mu\mathrm{s}$ and (b) $\tau=2000\,\mu\mathrm{s}$. Each point is classified from the sign structure of the measured processor energy change and the fluctuation relation bounds on heat and work as detailed in Table~\eqref{tab:thermodynamic-signs} of the Methods section. The solid boundary marks the transition between the observed thermodynamic modes, while the white boundary indicates an undefined region due to numerical precision. The shorter cycle shows a sharper boundary, whereas the longer cycle shifts and smooths the boundary, indicating that relaxation time changes the thermodynamic role of the same programmed instance. The experiment was performed the D-Wave Advantage6.4 system.}
\label{fig_1D_phase_diagram}
\end{figure*}
The choice of $\beta_1{=}1$ for the initial Gibbs state sets the absolute scale of the free energy bias between the prepared system and the environment but does not alter the infrared structure; hence it affects prefactors, not the $\log L$ form. Consistently, away from the critical region—where the correlation length is finite and the low–$k$ contribution is cut off by $1/\xi$ rather than $1/L$—the logarithmic growth is suppressed and per–spin quantities tend to flatten with $L$. Overall, the empirical $L\log L$ law is the expected finite size signature of critical fluctuations in one dimension under our protocol, and its appearance in TUR lower bounds reflects that these bounds are built from the first two cumulants of the energy change, which themselves are integrals of energy–energy correlations dominated by long wavelength modes. 

Figure~\eqref{fig_1D_phase_diagram} gives the cleanest view of how initialization and schedule depth select the operating mode in a one dimensional random uniform instance. The horizontal axis is the reverse annealing turning point \(s_{\rm p}\), which determines how far the cycle moves away from the classical problem Hamiltonian and into the region where the transverse driver is active. The vertical axis is the unitless initial inverse temperature \(\beta_1\), which sets the energy bias of the input ensemble. The two panels compare annealing times of \(\tau=100\,\mu\mathrm{s}\) and \(\tau=2000\,\mu\mathrm{s}\). In both cases the diagram separates primarily into two phases allowed by the TUR bounds: heater- and accelerator-compatible regions. Thus, for this instance and this range of parameters, the control field supplies positive work over the cycle; what changes across the boundary is whether the processor gains energy together with the environment, giving heater operation, or loses energy while the environment gains energy, giving accelerator operation.

Physically, at lower \(\beta_1\), the initial state is hotter in problem Hamiltonian units. The reverse annealing cycle then allows the processor distribution to relax towards lower problem energy, so that \(\langle \Delta E_1\rangle<0\). If the environment gains energy and the drive also supplies work, the device is an accelerator: it assists heat flow and relaxation rather than extracting work from them. At higher \(\beta_1\), the prepared ensemble is colder and has less excess problem energy to release. In this regime the same driven cycle can inject energy into both the processor and the environment, giving heater behaviour. The boundary therefore marks the point at which the sign of the processor energy change, and hence the thermodynamic role of the drive, switches.

\begin{figure*}
    \centering
    \includegraphics[width=1\linewidth]{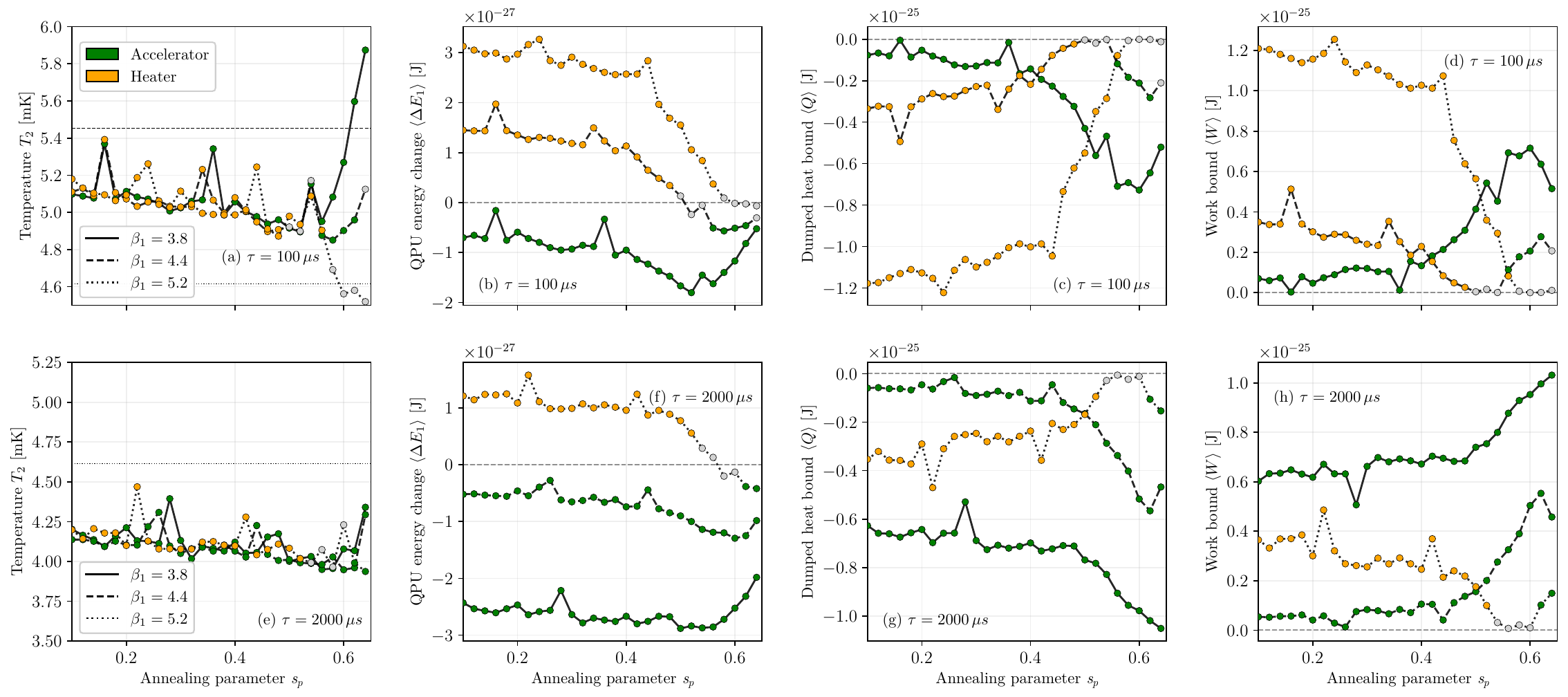}
     \caption{\textbf{Thermodynamic scans in a one dimensional random uniform Ising chain.}
A chain of \(N=300\) spins, with couplings \(J_{ij}\in[-1,1]\) and
longitudinal fields \(h_i\in[-h,h]\), is initialized from Gibbs ensembles at
\(\beta_1=3.8\), \(4.4\), and \(5.2\). The reverse annealing turning point
\(s_{\rm p}\) is scanned for two cycle durations: panels (a)--(d) show
\(\tau=100\,\mu\mathrm{s}\), and panels (e)--(h) show
\(\tau=2000\,\mu\mathrm{s}\). Panels (a) and (e) report the effective
temperature \(T_2\) inferred by pseudo-likelihood thermometry; the horizontal
reference lines indicate the temperatures associated with the initial
preparations. Panels (b) and (f) show the measured mean processor energy change
\(\langle\Delta E_1\rangle\). Panels (c) and (g) show the upper bound
 \(Q<0\) which allows
energy gain by the environment. Panels (d) and (h) show the lower bound on the work supplied during the cycle. Solid, dashed, and dotted
curves correspond to \(\beta_1=3.8\), \(4.4\), and \(5.2\), respectively,
while marker colours denote the assigned operating mode; grey markers indicate
points at which the sign criteria do not provide a unique assignment. The
\(\beta_1=3.8\) preparation is predominantly classified as an accelerator,
whereas \(\beta_1=5.2\) is predominantly classified as a heater. The
intermediate preparation follows the crossover between these regimes.
Increasing the cycle duration shifts and smooths this crossover while
preserving the same two operating modes. The experiments were performed on the D-Wave Advantage6.4 system.}
    \label{fig_1D_300}
\end{figure*}
When the annealer is classified as a \emph{heater}, the external schedule
supplies positive work and both the processor and its environment gain energy
over the cycle. Figure~\ref{fig_1D_300} connects this definition to the measured
and bounded thermodynamic quantities. For the colder preparation,
\(\beta_1=5.2\), the processor energy change is positive over most of the shallow
and intermediate turning point range for both
\(\tau=100\,\mu\mathrm{s}\) and \(\tau=2000\,\mu\mathrm{s}\)
[panels (b) and (f)]. At the same points, the upper bound on
\(Q=-\Delta E_2\) is negative [panels (c) and (g)], implying that the
environment gains energy, while the lower bound on work remains positive
[panels (d) and (h)], certifying work input by the control schedule. As the annealing parameter
\(s_{p}\) approaches the largest values in the scan, the energy change and
the heat and work bounds approach zero, and the distinction between the modes
becomes less pronounced. Heater operation is therefore the simplest driven
dissipative regime: energy supplied by the control fields is neither extracted
as work nor used to drive heat against the preparation bias, but is transferred
to the accessible processor and environmental degrees of freedom. When the
annealer is classified as an \emph{accelerator}, the processor loses energy
while the environment gains energy, but the control field still supplies
positive work. The warmer preparation, \(\beta_1=3.8\), displays this sign
structure over almost the full scan for both cycle times:
\(\langle\Delta E_1\rangle<0\), \(Q<0\), and
\(W>0\). The intermediate preparation, \(\beta_1=4.4\), illustrates
the effect of the cycle duration. It is predominantly heater-like at smaller
and intermediate \(s_{\rm p}\) for \(\tau=100\,\mu\mathrm{s}\), but
predominantly accelerator-like for \(\tau=2000\,\mu\mathrm{s}\). Thus, a longer
cycle shifts the crossover by allowing a different amount of relaxation,
without introducing a new class of operation. The effective temperature in
panel (e) is also lower and less variable than in panel (a), while the energy
and bound curves in panels (f)--(h) vary more smoothly.

The comparison between the two annealing times shows that increasing the annealing time from \(100\,\mu\mathrm{s}\) to \(2000\,\mu\mathrm{s}\) shifts and smooths the boundary, but it does not create a qualitatively different phase diagram. This shows that the annealing time mainly changes how much relaxation can occur during the cycle, and therefore where the sign change takes place. It does not change the underlying classification principle. This observation justifies using \(\tau=100\,\mu\mathrm{s}\) in the subsequent experiments: the shorter cycle gives a sharper finite time boundary and resolves the operating regimes more clearly, while retaining the same thermodynamic structure seen at longer times.
\paragraph*{Two dimensional instances.}
To test whether the observed energy exchange behaviour for 1D instances persists for more connected problems, we apply the same reverse annealing cycle and two-point energy change analysis to two representative two dimensional instances on a Pegasus P6 problem graph: a random uniform (RAU) instance, and a corrupted biased ferromagnet (CBFM)~\cite{pang2020potentialquantumannealingrapid}, chosen to span increasing landscape ruggedness. In CBFM, each qubit field is assigned as $h_i=-1$ with probability 0.85 and $h_i=0$ with probability 0.15. Each available coupler is assigned as $J_{ij}=+1$, $J_{ij}=-1$, or $J_{ij}=0$ with probabilities 0.55, 0.10, and 0.35, respectively. For each instance we scan the reverse annealing turning point $s_p$ for an annealing time of $\tau = 100\mu{\rm s}$.
From the observed statistics of the quantum processing unit energy change $\Delta E_1$ we extract (i) an effective environment temperature $T_2$ via pseudo-likelihood thermometry and (ii) TURs lower bounds on the average work and power per spin.
\begin{figure*}
\includegraphics[width=\textwidth]{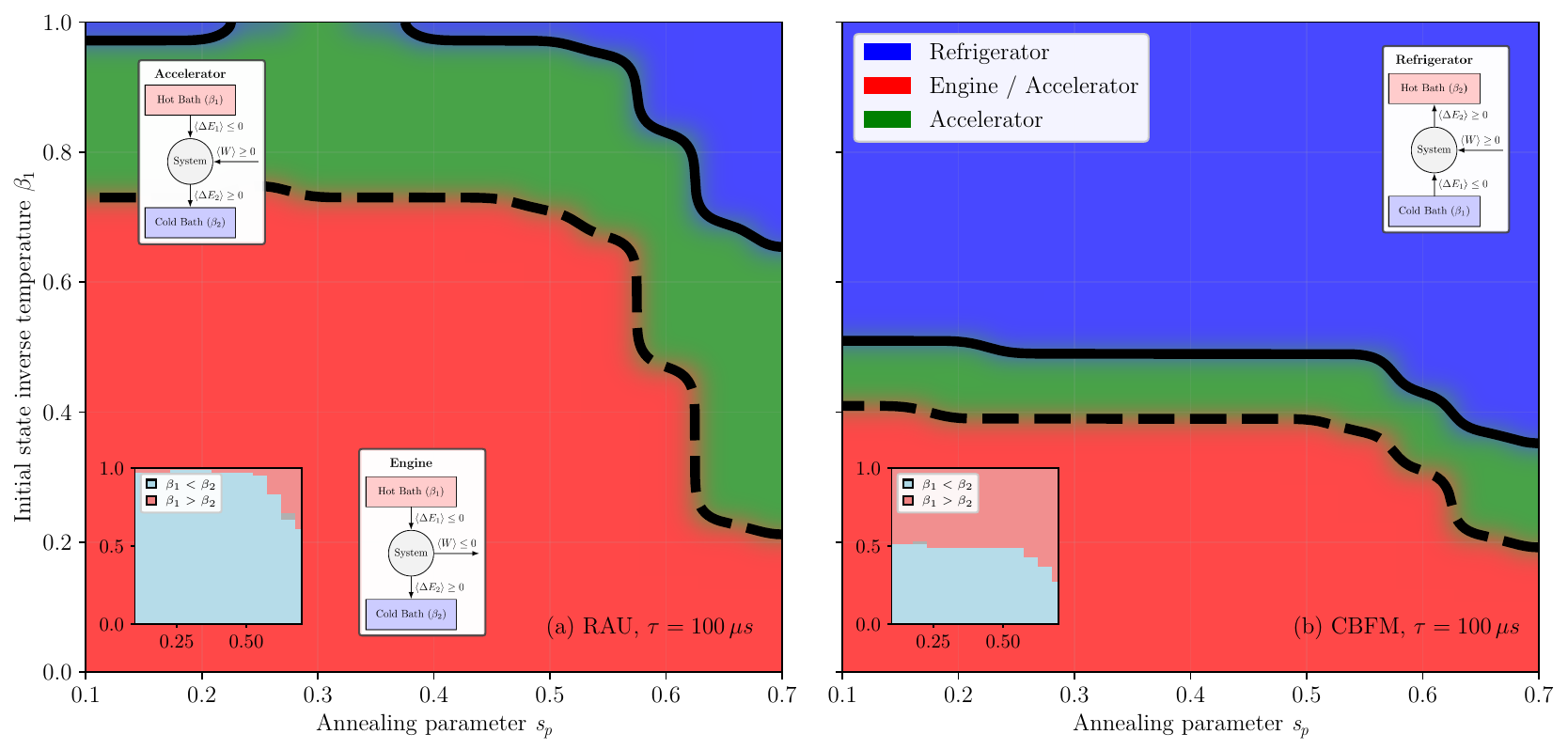}
     \caption{\textbf{Thermodynamic phase diagrams of two dimensional Pegasus instances.}
The operating mode is shown in the $(\beta_1,s_{\rm p})$ plane for a two dimensional RAU and CBFM Ising instance embedded natively on Pegasus and evolved for $\tau=100\,\mu\mathrm{s}$. Colours denote the thermodynamic label assigned from the measured processor energy change and the bounded heat and work signs. The phase boundary remains well defined, as in the one dimensional case, but its shape is modified by the higher connectivity and by the larger number of relaxation pathways available in the two dimensional problem. This demonstrates that the thermodynamic role of the annealer is controlled jointly by initialization, schedule depth, and problem structure. The experiments were performed on the D-Wave Advantage6.4 system.}
\label{fig_phase_diagram_pz}
\end{figure*}

Figure~\eqref{fig_phase_diagram_pz} extends the operating mode analysis to the two dimensional RAU and CBFM instance at \(\tau=100\,\mu\mathrm{s}\), where we see that the same experimentally accessible controls, \(\beta_1\) and \(s_{\rm p}\), maps the TUR-allowed thermodynamic regions. The boundary is no longer identical to that of the one dimensional instance, but this is expected: increasing the connectivity changes the spectrum of accessible configurations, the number of relaxation pathways, and the probability that defects are created or removed during the cycle. The engine/accelerator-compatible label is retained deliberately. Since only
\(\Delta E_1\) is measured directly, the signs of heat and work are inferred through
bounds. In regions where the available bounds do not uniquely separate work
extraction from driven acceleration, assigning a combined label avoids overinterpreting
the data. Therefore, this shows that the thermodynamic role of the annealer is programmable but not universal. It is programmable because changing \(\beta_1\) or \(s_{\rm p}\) can move the same hardware between refrigerator, and engine/accelerator thermal operation allowed by TURs. It is not universal because the position and shape of the boundaries depend on the problem graph. In a two dimensional instance, each spin is coupled to more neighbours than in a chain, so a local change can affect more terms in the problem Hamiltonian. This increases the number of possible relaxation routes and changes the balance between energy lowering, excitation production, and heat exchange with the environment. The resulting boundary is therefore a property of the full experimental protocol: hardware, problem instance, initialization, and schedule.
\begin{figure}[t!]
    \centering
    \includegraphics[width=\textwidth]{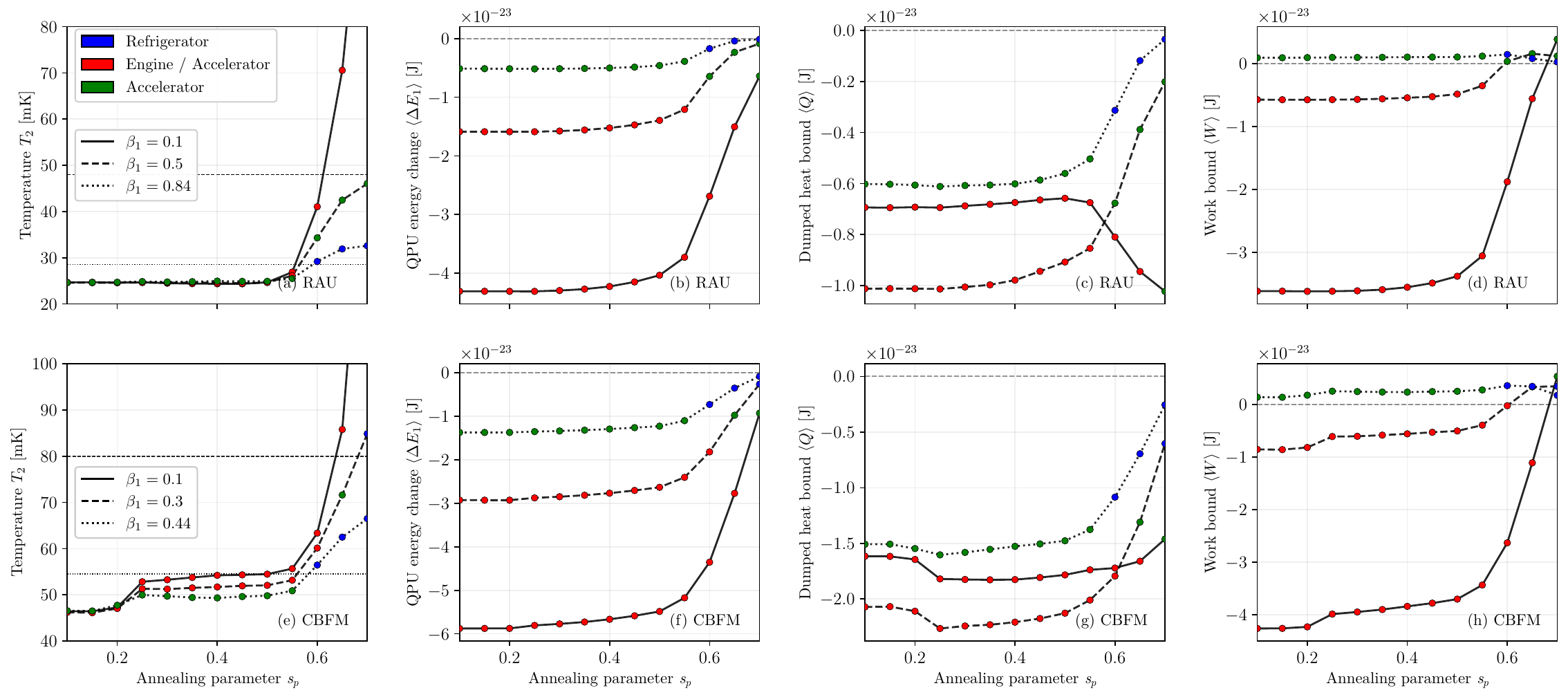}
    \caption{\textbf{Scans underlying the thermodynamic classification at $\tau=100\,\mu\mathrm{s}$.}
The reverse annealing turning point $s_{\rm p}$ is scanned for several preparation temperatures. Panels (a--d) show the random uniform instance and panels (e--h) show the corrupted biased ferromagnet (CBFM) used as a more rugged comparison. The plotted quantities are the effective environment temperature $T_2$, the mean processor energy change $\langle\Delta E_1\rangle$, the bound on heat dumped into the environment, and the bound on work. The rapid changes of all quantities in the same interval of $s_{\rm p}$ identify the active window that controls the phase boundaries. The comparison with the corrupted biased ferromagnet shows that problem structure broadens this window and increases the baseline thermodynamic cost. The experiments were performed on the D-Wave Advantage6.4 system.}
\label{fig_2d_scans}
\end{figure}

Figure~\eqref{fig_2d_scans} displays how the effective temperature \(T_2\), the mean processor energy change \(\langle\Delta E_1\rangle\), the heat bound, and the work bound vary as \(s_{\rm p}\) is scanned at fixed \(\tau=100\,\mu\mathrm{s}\). The most important feature is that these quantities change together. The phase boundary is therefore not an artifact of a single fitted parameter. It is supported simultaneously by the drift of the processor energy distribution and by the fluctuations that enter the thermodynamic bounds.
For the random uniform instance, the response is relatively sharp. Over much of the scan the annealer remains in a weakly perturbed regime: the output distribution changes slowly with \(s_{\rm p}\), the inferred temperature\(T_2\) is nearly constant, and the heat and work bounds remain modest. In a narrower interval of \(s_{\rm p}\), the mean energy change and the thermodynamic bounds vary rapidly. This is the schedule region where the reverse annealing cycle most efficiently changes the output distribution. Physically, the system is then deep enough into the schedule to allow transitions between nearby configurations, but not so deep or so shallow that the dynamics become either fully randomized or almost frozen. In this window, small changes in \(s_{\rm p}\) produce large changes in relaxation and excitation probabilities, which is why the operating mode can change over a narrow range of the control parameter.

The comparison with the corrupted biased ferromagnet shows that the thermodynamic response is instance dependent. The corrupted biased ferromagnet has a more structured energy landscape, with local defects and competing tendencies that are harder to remove during the return branch of the cycle. As a result, the scans are broader and the baseline energy changes are larger. This means that the annealer spends a wider range of \(s_{\rm p}\) values in a thermodynamically active regime. In practical terms, the same reverse annealing schedule carries a different energetic cost depending on the problem. A schedule that is nearly reversible for a smoother random uniform instance may generate larger heat and work bounds for a more rugged instance. Figure~\eqref{fig_2d_scans} therefore provides the microscopic explanation for the phase diagrams: the boundaries arise where the measured energy drift and its fluctuations become large enough to change the inferred signs of heat and work.

Because the operating mode maps depend on the prescribed initial Gibbs
ensemble, we separately examined the convergence of the classical sampler used
for the RAU and CBFM preparations. The sampled energy and magnetization reach
stationary plateaus over the production burn-in range, with magnetization close
to zero for RAU and finite magnetization for CBFM, as expected from their
different coupling structures. The corresponding diagnostics are reported in
Appendix~\eqref{app:gibbs-quality}. These checks support stationarity of the
low order observables that determine the initial energy scale, while not
establishing that the complete high dimensional distribution is exactly
Gibbsian. To examine the dependence on hardware generation, we repeated the RAU and
CBFM analysis on a D-Wave Advantage2 processor using the same thermodynamic prescription. Appendix~\eqref{app:advantage2-results} shows that the same ordering
of engine/accelerator-compatible, and refrigerator regions is
recovered, although the fitted effective temperatures and the locations of the
mode boundaries differ between instances and processors. This comparison
supports transferability of the classification procedure, but not universality
of the estimated temperatures or operating mode boundaries.

In summary, we showed that reverse annealing can be used as a thermodynamic control protocol. The D-Wave processor is not merely sampled after an anneal; it is driven through a closed cycle whose energy balance can be classified. Heater-, accelerator-, engine-, and refrigerator-compatible operations are therefore not metaphors for computational performance. They are experimentally defined energy flow regimes of the programmed quantum processing unit degrees of freedom. This provides a useful layer of characterization in addition to success probability or time to solution: it identifies which schedules solve or sample a problem with small thermodynamic cost, which schedules deliberately enhance relaxation, and which schedules cross into regimes where work and heat change sign.

\section{Discussion}
We studied a commercial quantum annealer as a programmable thermal machine whose operating mode can be selected by experimentally accessible controls. Building on reverse annealing cycles and two-point energy change measurements, we demonstrated that by tailoring the schedule and the initialization temperature the same device can realize multiple thermodynamic behaviours allowed by the TUR bounds, and that these regimes can be diagnosed operationally from the sign structure of the cycle averaged energy exchanges without invoking a detailed microscopic bath model. 

A central technical outcome is that TURs, combined with the exchange fluctuation framework, turn the limited energy readout available on today’s annealers into information about thermodynamic performance. Concretely, the first two moments of the measured processor energy change $\Delta E_1$ suffice to place bounds on entropy production and on per-cycle heat and work, and hence to obtain bounds on per-cycle power for any chosen protocol. This methodology is device agnostic and relies only on statistics that are already accessible at the user level, enabling thermodynamic benchmarking to be carried out alongside standard computational metrics.

Across one and two dimensional instances, we found that energy exchange and irreversibility concentrate within a relatively narrow mid-anneal window where the driver and problem energy scales compete most strongly. In this window, effective thermometry signatures and TUR-based dissipation bounds become most pronounced, while outside it the TUR-based lower bounds on dissipation are smaller, consistent with millikelvin operation. The resulting operating mode phase diagrams in the $(\beta_1,s_p)$ plane provide a compact, experimentally grounded map of when the TUR-allowed heater, engine, accelerator and refrigerator behaviour is accessible.

A connection between the thermodynamic classification and computation can be established by the measured processor energy change \(\Delta E_1\). When an optimization problem is
encoded consistently in the Ising energy \(E_{\rm p}\), including sufficiently
strong penalties for constraints, \(\Delta E_1<0\) means that the final
bitstring has a lower programmed cost than the initial bitstring, whereas
\(\Delta E_1>0\) means that its programmed cost has increased. In the
one-dimensional accelerator regime observed here, the processor loses energy
while the control schedule supplies positive work. Computationally, this
corresponds to a driven refinement step in which external control assists the
transition from the supplied initial candidate toward configurations with
lower objective energy. Accelerator operation does not guarantee that the
global optimum is reached, but it shows that the protocol improves the encoded
objective on average relative to its starting ensemble. Heater operation has
the opposite final state effect: the processor gains energy and the returned
configurations have a higher programmed cost on average. It is therefore
generally undesirable as the final stage of an optimization run. A controlled
heater-like stage could nevertheless be tested as an intermediate
diversification step that moves the system away from a local basin before a
later refinement cycle. The present measurements do not establish such a
computational benefit, but reverse annealing and pausing studies show that
changing the accessible search range and relaxation time can substantially
alter the probability of finding improved solutions
\cite{Chancellor2021SearchRange,Vrinda2025,Pausing2020,
Pausing2022,Smierzchalski_2024}.

The refrigerator- and engine-compatible phase require a more careful
computational interpretation. Refrigerator operation means that the drive
forces energy exchange against the spontaneous thermal bias. Depending on
whether \(\beta_1<\beta_2\) or \(\beta_1>\beta_2\), this operation can be
associated with either a decrease or an increase of the processor energy.
Consequently, refrigerator operation represents computational improvement
only at points where \(\Delta E_1<0\); the thermodynamic footprint alone is not
sufficient. Likewise, an engine-compatible point should not be interpreted as
a computational speedup or as a solution obtained without cost. In the
present analysis, the work bound does not uniquely distinguish
work extraction from driven acceleration, which is why the combined
engine/accelerator-compatible label is retained. Computational performance
must therefore be assessed separately through the probability of improving
the initial state, the fraction of feasible outputs, the probability of
reaching a target energy, the mean and best objective gaps, sample diversity,
and time to solution. A direct extension of the present work would report
these quantities conditionally on the thermodynamic mode. Such an analysis
would determine whether particular modes predict useful refinement,
exploration, or loss of solution quality, rather than assuming this relation
from thermodynamic signs alone
\cite{DoucetNJP,Smierzchalski_2024,HanussekPRAp2026,
jakub_PRAp2026,TuzPRAp2026}.

The thermometry analysis provides an additional operational interpretation of
these phase diagrams. The inverse temperature \(\beta_2\) obtained from
pseudo-likelihood fitting should be understood as an effective temperature
parameter of the sampled Ising degrees of freedom, rather than as a direct
measurement of the dilution refrigerator plate temperature. It results from the
physical temperature with the programmed energy scale, schedule dependent
relaxation, freeze-out, and possible deviations of the
output distribution from an exact Gibbs state. The simultaneous variation of
\(T_2\), \(\langle\Delta E_1\rangle\), and the TUR-based bounds within the
same mid-anneal interval therefore indicates that the estimator is sensitive to
the part of the schedule in which the sampled distribution is reorganized most
strongly. In addition, the line \(\beta_1=\beta_2\) separates the two possible
effective temperature orderings. Crossing this line reverses the spontaneous
direction of heat exchange and consequently changes which mixed sign
energy flow pattern corresponds to refrigerator or engine/accelerator
operation. Repeating the protocol for a fixed calibrated instance, embedding,
energy normalization, and schedule can therefore provide a built-in effective
thermometry diagnostic for the active qubits, with systematic shifts in
\(T_2\) serving as indicators of changes in relaxation, freeze-out, or
calibration. This diagnostic is complementary to, but does not replace,
cryogenic thermometry
\cite{Benedetti2016EffectiveTemperature,
Raymond2016GlobalWarming,Nelson2022HighQualityGibbs,
Grattan2025ClassicalThermometry}.

Our results are also relevant when the annealer is used as a probabilistic
sampler rather than as an optimizer. In this setting, the objective is not
necessarily to return one minimum energy configuration, but to produce a
distribution with a controlled balance between low energy weight and sample
diversity. Under an approximate Gibbs description, a lower effective
temperature concentrates probability on low energy configurations, whereas a
higher effective temperature broadens the distribution over a larger part of
the configuration space. The thermodynamic mode provides complementary
information about how this distribution was produced. Accelerator-like
relaxation can increase the weight of lower energy samples, while heater-like
operation broadens the sampled energy range; refrigerator operation indicates
that the control protocol drives the distribution against its spontaneous
thermal tendency. These distinctions can inform schedule selection in
Boltzmann machine training, probabilistic inference, and generative sampling,
where both the effective temperature and the diversity of the returned
configurations matter. The usefulness of a given mode for sampling must,
however, be evaluated together with a fitting test for the Gibbs
description, because a fitted temperature alone does not establish that the
complete output distribution is thermal
\cite{Benedetti2016EffectiveTemperature,
Nelson2022HighQualityGibbs,Amin_2015}.

The operating mode maps also have implications beyond optimization. In
statistical physics and materials simulations, quantum annealers are used to
study magnetic ordering, defect formation, phase transitions, and
nonequilibrium relaxation. For such applications, the distinction between
accelerator, heater, and refrigerator operation indicates whether the chosen
schedule mainly supports relaxation toward lower energy configurations,
broadens the sampled ensemble through net heating, or drives energy against an
effective thermal bias. This information can help separate properties of the
programmed model from artefacts introduced by the control schedule and
hardware environment. The effective thermometry and energy flow
classification are therefore relevant to simulations in which temperature,
thermalization, and relaxation rates are physical observables rather than
unwanted disturbances
\cite{Bando2020Universality,Sathe2026ClassicalCriticality,
King2025BeyondClassical}.

From a computer engineering perspective, the present bounds provide
chip-level information that is absent from conventional benchmarks based only
on runtime and solution quality. They can be used to compare schedules that
produce similar computational outputs but differ in their minimum
energy exchange. This comparison is only one part of a complete energy
assessment: programming, readout, control electronics, refrigeration,
embedding overhead, total runtime, and the probability of obtaining an
acceptable solution must also be included. Combining these levels would make
it possible to compare algorithms through energy to solution, rather than
through energy per annealing cycle alone
\cite{fellous2023,Meier2025,Smierzchalski_2024}.

The present results motivate four related directions. First, computational
metrics should be evaluated within each thermodynamic mode by measuring
objective improvement, feasibility, target-energy probability, sample
diversity, and time to solution. Second, schedule parameters such as cycle
time, pauses, and locally slowed ramps can then be selected jointly for
solution quality, sampling behaviour, and thermodynamic cost. Third, the same
classification can be applied to statistical-physics simulations to identify
where measured observables are dominated by relaxation, net heating, or
driven heat transfer. Fourth, the chip-level bounds should be integrated with
programming, control, readout, and refrigeration costs to obtain an
application-level energy-to-solution measure
\cite{Pausing2020,Pausing2022,Smierzchalski_2024,
Sathe2026ClassicalCriticality,fellous2023,Meier2025}.
Overall, the results show that a quantum annealer is not only an optimizer or
sampler, but a controlled computing system whose output distribution and
energy exchange respond jointly to the problem encoding, initialization, and
schedule. The thermodynamic classification therefore provides an additional
tool for selecting and comparing computational protocols, while remaining
complementary to direct measures of accuracy, runtime, and sampling quality.

\section{Methods}
We now detail the experimental and analytical procedures that underpin our results. Our methodology is modular and device agnostic: it specifies (i) the effective Hamiltonian realized by the quantum annealer, (ii) the reverse annealing protocol that implements closed thermodynamic cycles, and (iii) a measurement scheme based solely on two–point energy differences of the problem Hamiltonian. Building on this, we extract heat and work from energy–change statistics using exact fluctuation identities, and derive TUR lower bounds. The same framework allows for an operational classification of engine-,heater-, refrigerator-, and accelerator-compatible regimes from the sign structure of average energy exchanges, and provides an estimate on the power at finite annealing time. Throughout, we state assumptions explicitly (reverse annealing, two temperature preparation, and factorized initial state) and report sampling procedures and uncertainty quantification to ensure reproducibility.
\subsection{Hardware and model}
All demonstrations were performed on commercial D\textendash Wave quantum annealer (Advantage and Advantage2 systems) operated via the Leap cloud interface. The effective device Hamiltonian is a transverse--field Ising model.
\begin{equation}
\mathcal{H}_{\text{Ising}}(s)
= -\frac{A(s)}{2} \sum_i \sigma_i^{x}
  + \frac{B(s)}{2} \left(
      \sum_i h_i \sigma_i^{z}
      + \sum_{i>j} J_{ij} \sigma_i^{z} \sigma_j^{z}
    \right),
\qquad s \in [0,1].
\label{eq:H_of_s}
\end{equation}
with $A(s)$ and $B(s)$ being the transverse field (driver) schedule and problem Hamiltonian schedule, respectively. Pauli operators $\sigma_i^{x,z}$ on physical qubits, programmable fields $\{h_i\}$ and couplers $\{J_{ij}\}$, and a dimensionless annealing parameter $s\in[0,1]$ that sets the relative weights of driver and problem terms. Device--level parameters (qubit temperature, programming/readout times, slope limits) follow the vendor specifications and are held fixed throughout. 

The measurements were performed on three D-Wave chips whose main
specifications are summarized in Table~\eqref{tab:qpu-specifications}. The
Advantage processors implement the Pegasus P16 topology, for which a qubit
has a nominal maximum degree of 15, whereas the Advantage2 processor
implements the Zephyr Z12 topology with a nominal maximum degree of 20
\cite{DWavePegasus2019,DWaveZephyr2021,
DWaveAdvantage2Performance2025}. The increased Zephyr connectivity can
reduce the embedding overhead for non-native problem graphs, but it does not
change the thermometry procedure.
\begin{table}[h!]
\centering
\caption{
\textbf{Specifications of the D-Wave quantum processors}
}
\label{tab:qpu-specifications}
\small
\setlength{\tabcolsep}{4pt}
\begin{tabular}{lcccc}
\toprule
Solver identifier
&
Native topology
&
Maximum degree
&
Active qubits
&
Active couplers
\\
\midrule
\texttt{Advantage\_system5.4}
&
Pegasus P16
&
15
&
5614
&
40050
\\
\texttt{Advantage\_system6.4}
&
Pegasus P16
&
15
&
5612
&
40088
\\
\texttt{Advantage2\_system2.1}
&
Zephyr Z12
&
20
&
4516
&
40448
\\
\bottomrule
\end{tabular}
\end{table}
\subsection{Annealing protocols}
We implement reverse annealing cycles in which $s(t)$ starts and ends at $1$, dips to a programmable minimum $s_p$, then returns to $1$:
\begin{equation}
s(t)=
\begin{cases}
1-2(1-s_p)\,t/\tau, & t\in[0,\tau/2],\\[2pt]
-1+2 s_p+2(1-s_p)\,t/\tau, & t\in[\tau/2,\tau],
\end{cases}
\label{eq:rev_schedule}
\end{equation}
with total cycle time $\tau$. This cyclic schedule ensures $H(0)=H(1)\equiv H_p=\sum_i h_i\sigma_i^z+\sum_{\langle i,j\rangle}J_{ij}\sigma_i^z\sigma_j^z$, which is crucial for the two--point energy measurement used below.
\subsection{State preparation}
Each run begins from a classical spin configuration $\boldsymbol{\sigma}\in\{\pm1\}^N$ sampled from the Boltzmann distribution of $H_p$ at an effective inverse temperature $\beta_1$:
\begin{equation}
\mathbb{P}\!\left[\sigma\right]=\frac{e^{-\beta_1 E_p(\boldsymbol{\sigma})}}{Z(\beta_1)},\qquad
E_p(\boldsymbol{\sigma})=\sum_i h_i\sigma_i+\sum_{\langle i,j\rangle}J_{ij}\sigma_i\sigma_j .
\label{eq:hot_init}
\end{equation}
The processor remains coupled throughout the cycle to its cryogenic
environment, represented within the effective two temperature description by
the inverse temperature parameter \(\beta_2\). In practice, \(\beta_2\) is
estimated from pseudo-likelihood fits to the output configurations, while
\(\beta_1\) is fixed by the classical sampling routine used to prepare the
initial configurations. No fixed ordering between \(\beta_1\) and
\(\beta_2\) is imposed across the complete parameter scan; their relative
ordering is determined point by point and identifies which subsystem is
effectively hotter.

\subsection{Measured observable and two--point energy change}
Because $H(0)=H(1)=H_p$, the processor energy change over one reverse cycle is accessible by two--point measurement of $H_p$:
\begin{equation}
\Delta E_1=E_p\!\bigl(\boldsymbol{\sigma}(\tau)\bigr)-E_p\!\bigl(\boldsymbol{\sigma}(0)\bigr),
\label{eq:dE1}
\end{equation}
where $\boldsymbol{\sigma}(\tau)$ is the readout configuration at $t=\tau$. Repeating the cycle yields the empirical distribution $p(\Delta E_1)$ and its first two moments $\langle \Delta E_1\rangle$ and $\langle \Delta E_1^2\rangle$. We collect at least $10^4$ cycles per setting to suppress statistical error and to tighten the bounds described below. \hfill {\small \,\!}

\subsection{Exchange fluctuation theorem and entropy production}
Consider the compound (system + environment) initialized in the factorized thermal state
\begin{equation}
\rho_0=\frac{e^{-\beta_1 H_p}}{Z_p(\beta_1)}\otimes\frac{e^{-\beta_2 H_E}}{Z_E(\beta_2)}\, .
\label{eq:fact_state}
\end{equation}
For any cyclic protocol ($H(0)=H(1)=H_p$), the joint distribution of the stochastic energy changes $\Delta E_{1,2}$ of the processor and the environment obeys the exchange fluctuation theorem
\begin{equation}
\frac{p(\Delta E_1,\Delta E_2)}{p(-\Delta E_1,-\Delta E_2)}=\exp\!\left(\beta_1 \Delta E_1+\beta_2 \Delta E_2\right),
\label{eq:xFT}
\end{equation}
which implies the second law in the form $\langle \Sigma\rangle\equiv\beta_1\langle \Delta E_1\rangle+\beta_2\langle \Delta E_2\rangle\ge 0$~\cite{Jarzynski1997, Crooks1999, campisi2011, esposito2009}. Here we identify average heat dumped into the bath as $\langle Q\rangle\equiv -\langle \Delta E_2\rangle$ and average driving work as $\langle W\rangle\equiv \langle \Delta E_1\rangle+\langle \Delta E_2\rangle$.

\subsection{TUR--based bounds from $\Delta E_1$ alone}
Only $\Delta E_1$ is directly accessible on current hardware. Using \eqref{eq:xFT}, the thermodynamic uncertainty relations (TURs) bound entropy production and, in turn, heat and work in terms of the first two moments of $\Delta E_1$ alone. Defining $g(x)=x\,\mathrm{artanh}(x)$, one finds~\cite{Barato2015, Gingrich2016, Horowitz2020, Hasegawa2019}
\begin{align}
\langle \Sigma\rangle&\ge 2\,g\!\left(\frac{\langle \Delta E_1\rangle}{\sqrt{\langle \Delta E_1^2\rangle}}\right), \label{eq:TURsigma}\\[2pt]
-\langle Q\rangle&\ge \frac{2}{\beta_2}\,g\!\left(\frac{\langle \Delta E_1\rangle}{\sqrt{\langle \Delta E_1^2\rangle}}\right)-\frac{\beta_1}{\beta_2}\,\langle \Delta E_1\rangle, \label{eq:TURQ}\\[2pt]
\langle W\rangle&\ge \frac{2}{\beta_2}\,g\!\left(\frac{\langle \Delta E_1\rangle}{\sqrt{\langle \Delta E_1^2\rangle}}\right)+\Bigl(1-\frac{\beta_1}{\beta_2}\Bigr)\,\langle \Delta E_1\rangle.
\label{eq:TURW}
\end{align}
Equations~\eqref{eq:TURQ}--\eqref{eq:TURW} provide device--certified, bath model--free lower bounds on $-\langle Q\rangle$ and $\langle W\rangle$ using only $\beta_{1,2}$ and energy--change statistics. The inverse temperature $\beta_2$ of the environment is obtained by fitting the sampled spin configurations to the Boltzmann distribution encoded by the programmed couplings and fields~\cite{Campisi_2021, pseudo1, pseudo2}. Concretely, for a dataset ($\mathcal{D}
=\{\boldsymbol{s}^{(1)},\ldots,\boldsymbol{s}^{(D)}\}$), the average pseudo-likelihood ($\Lambda(\beta)$) is evaluated as a function of $\beta_2$, and given by:
\begin{equation}
    \Lambda(\beta) = -\frac{1}{ND} \sum_{i=1}^N \sum_{d=1}^D \ln\left[ 1 + \exp \left( -2 \beta s_i^{(d)} \Bigl( h_i + \sum_{j \in \delta_i} J_{ij} s_j^{(d)} \Bigr) \right) \right].
\end{equation}
The fitted inverse temperature maximizes the average pseudo-likelihood. Thus
\begin{equation}
    \beta_2 = \arg\max_{\beta} \Lambda(\beta)
    \label{beta2_est}
\end{equation}
states that $\beta_2$ is chosen as the parameter that makes the observed samples most consistent, spin by spin, with the conditional probabilities implied by the Ising model actually run on the device. 

The pseudo-likelihood procedure returns a dimensionless inverse temperature
parameter \(\beta_2\), because the fit is performed using the dimensionless
classical Ising energy.
To express this parameter as a temperature, we use the chip specific annealing
functions supplied by D-Wave
\cite{DWaveAnnealingControls2026,DWavePerQPUSchedules2026,
DWaveFreezeoutTemperature2026}.
The physical energy associated with a classical configuration at an annealing
fraction $s$ is

\begin{equation}
\mathcal{E}_{\rm p}
\left(s_p,\boldsymbol{\sigma}\right)
=
\frac{h_{\rm P}\,10^{9}}{2}
B_{\rm GHz}\left(s_p\right)
E_{\rm p}(\boldsymbol{\sigma}),
\label{eq:physical-ising-energy}
\end{equation}
where \(h_{\rm P}\) is Planck's constant and
\(B_{\rm GHz}(s)\) denotes the numerical value of the D-Wave
problem Hamiltonian schedule in GHz. The subscript \({\rm P}\) on
\(h_{\rm P}\) distinguishes Planck's constant from the programmed local fields
\(h_i\).

Matching the fitted distribution
\begin{equation}
p(\boldsymbol{\sigma})
\propto
\exp\left[-\beta_2 E_{\rm p}(\boldsymbol{\sigma})\right]
\end{equation}
to the physical Boltzmann distribution
\begin{equation}
p(\boldsymbol{\sigma})
\propto
\exp\left[
-\frac{
\mathcal{E}_{\rm p}(s_p,\boldsymbol{\sigma})
}{
k_{\rm B}T_2
}
\right]
\end{equation}
gives
\begin{equation}
\beta_2(s_p)
=
\frac{
h_{\rm P}\,10^{9}
B_{\rm GHz}(s_p)
}{
2k_{\rm B}T_2(s_p)
}.
\label{eq:beta-physical-temperature}
\end{equation}
The effective temperature in millikelvin is therefore
\begin{align}
T_2(s_p)\,[{\rm mK}]
&=
10^3
\frac{
h_{\rm P}\,10^{9}
B_{\rm GHz}(s_p)
}{
2k_{\rm B}\beta_2(s_p)
}
\nonumber\\
&=
23.9962\,
\frac{
B_{\rm GHz}(s_p)
}{
\beta_2(s_p)
}.
\label{eq:temperature-mK}
\end{align}
The numerical coefficient in Eq.~\eqref{eq:temperature-mK} follows from
\(h_{\rm P}=6.62607015\times10^{-34}\,{\rm J\,s}\) and
\(k_{\rm B}=1.380649\times10^{-23}\,{\rm J\,K^{-1}}\). The factor of
\(1/2\) follows from the D-Wave Hamiltonian convention used above.

Combining \(\langle\Sigma\rangle\geq 0\) with
\(\langle W\rangle=
\langle\Delta E_1\rangle+\langle\Delta E_2\rangle\)
constrains the thermodynamically allowed sign patterns. Since
\(\beta_1\) is controlled by the preparation procedure, whereas
\(\beta_2\) is inferred from the output ensemble, we do not impose a fixed
ordering between them. Both \(0<\beta_1<\beta_2\) and
\(0<\beta_2<\beta_1\) can occur across the parameter scans. The ordering
determines which subsystem is effectively hotter and therefore fixes the
spontaneous direction of energy exchange. Reversing the ordering interchanges
the roles of \(\Delta E_1\) and \(\Delta E_2\) in the refrigerator,
engine, and accelerator modes, while the heater sign pattern remains
unchanged. Table~\eqref{tab:thermodynamic-signs} summarizes the thermodynamically allowed sign conventions between the processor energy change and its environment.
\begin{table}[h!]
\centering
\caption{
Thermodynamic sign conventions for the processor energy change
\(\langle\Delta E_1\rangle\), environment energy change
\(\langle\Delta E_2\rangle\), and driving work
\(\langle W\rangle=\langle\Delta E_1\rangle+\langle\Delta E_2\rangle\).
Positive (negative) work denotes energy supplied by (extracted from) the drive.
}
\label{tab:thermodynamic-signs}
\begin{tabular}{lccc}
\hline
\multicolumn{4}{c}{
\(0<\beta_1<\beta_2\)
\quad (\(T_1>T_2\): subsystem \(1\) is hotter)
}
\\
\hline
Operation
&
\(\langle\Delta E_1\rangle\)
&
\(\langle\Delta E_2\rangle\)
&
\(\langle W\rangle\)
\\
\hline
Refrigerator [R] & \(\geq 0\) & \(\leq 0\) & \(\geq 0\) \\
Engine [E]       & \(\leq 0\) & \(\geq 0\) & \(\leq 0\) \\
Accelerator [A]  & \(\leq 0\) & \(\geq 0\) & \(\geq 0\) \\
Heater [H]       & \(\geq 0\) & \(\geq 0\) & \(\geq 0\) \\
\hline
\multicolumn{4}{c}{
\(0<\beta_2<\beta_1\)
\quad (\(T_2>T_1\): subsystem \(2\) is hotter)
}
\\
\hline
Operation
&
\(\langle\Delta E_1\rangle\)
&
\(\langle\Delta E_2\rangle\)
&
\(\langle W\rangle\)
\\
\hline
Refrigerator [R] & \(\leq 0\) & \(\geq 0\) & \(\geq 0\) \\
Engine [E]       & \(\geq 0\) & \(\leq 0\) & \(\leq 0\) \\
Accelerator [A]  & \(\geq 0\) & \(\leq 0\) & \(\geq 0\) \\
Heater [H]       & \(\geq 0\) & \(\geq 0\) & \(\geq 0\) \\
\hline
\end{tabular}
\end{table}

Given the annealing time $\tau$, we report a bound on the power
\begin{equation}
P_{\mathrm{bound}} \ge \frac{W}{\tau},\qquad 
W\equiv \frac{2}{\beta_2}\,g\!\left(\frac{\langle \Delta E_1\rangle}{\sqrt{\langle \Delta E_1^2\rangle}}\right)+\Bigl(1-\frac{\beta_1}{\beta_2}\Bigr)\langle \Delta E_1\rangle.
\label{eq:power}
\end{equation}

\subsection{Experimental procedure and statistics}
For each annealing parameter and annealing duration $(s_p, \tau)$ we: (i) draw $M$ initial configurations from \eqref{eq:hot_init}; (ii) execute a reverse cycle \eqref{eq:rev_schedule} and read out $\sigma(\tau)$; (iii) compute $\Delta E_1$ via \eqref{eq:dE1}; (iv) accumulate moments $\langle \Delta E_1\rangle$, $\langle \Delta E_1^2\rangle$; (v) estimate $\beta_2$ via pseudo-likelihood; (vi) evaluate the bounds \eqref{eq:TURsigma}; and (vii) assign the thermodynamic regime using the sign structure above. We use $M\!\ge\!10^4$ cycles per point (unless otherwise stated).

\subsection{Assumptions, controls, and reproducibility}
Our analysis assumes: (i) cyclic schedule $H(0)=H(1)$; (ii) initial factorized thermal state \eqref{eq:fact_state}; and (iii) validity of the exchange fluctuation theorem \eqref{eq:xFT} for the compound. TUR bounds \eqref{eq:TURsigma}--\eqref{eq:TURW} follow directly from \eqref{eq:xFT} and do not require Markovianity or weak coupling, making them robust to device nonidealities. We verified that conclusions are stable under moderate changes of $s_p$, under optional pauses, and across programming batches.

\appendix
\section{Validation of the prepared Gibbs ensembles}
\label{app:gibbs-quality}
\begin{figure}[th!]
\centering
\includegraphics[width=0.95\linewidth]{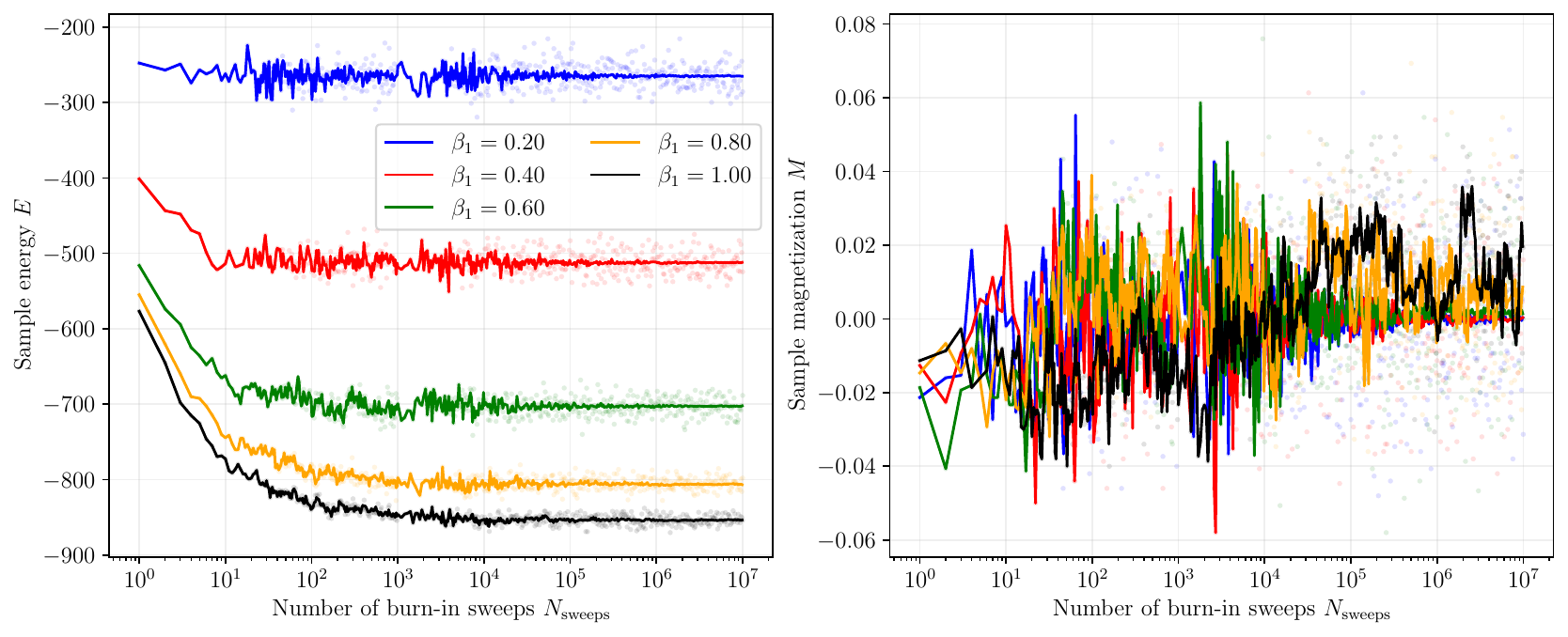}\\
\includegraphics[width=0.95\linewidth]{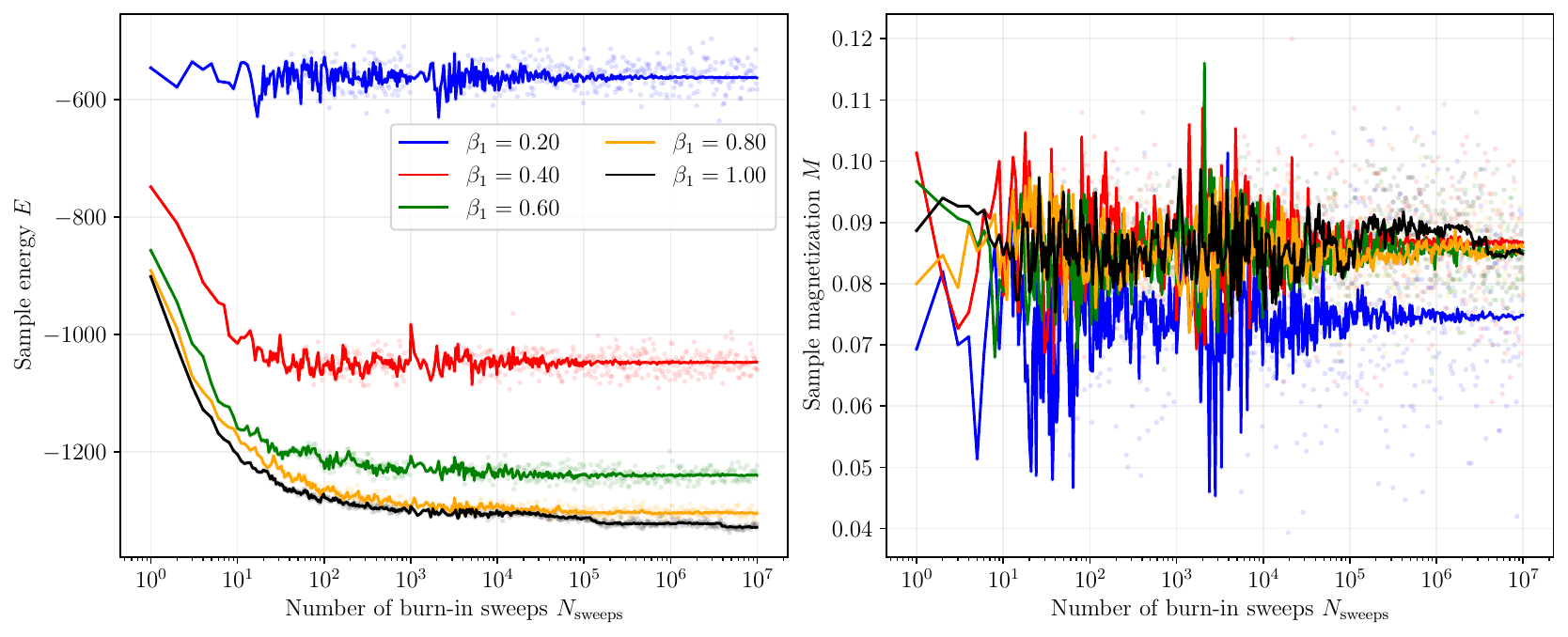}
\caption{
\textbf{Burn-in diagnostics for the Gibbs initial samples.}
Sample energy \(E\) and magnetization \(M\) are shown as functions of the number
of burn-in sweeps for the RAU (upper row) and CBFM (lower row) Pegasus P6 instances. In both cases the
energy reaches a plateau after an initial relaxation period. The RAU magnetization
fluctuates around zero, while the CBFM magnetization remains finite because of the
biased ferromagnetic structure of the instance. The plateau region is used to choose
the burn-in length for the production samples.
}
\label{fig:gibbs-burnin}
\end{figure}
The thermodynamic analysis assumes that the initial configurations are sampled from
a Gibbs distribution of the programmed problem Hamiltonian at inverse temperature
\(\beta_1\). We therefore checked the convergence of the classical sampler used to
prepare the input states before sending them to the quantum annealer. Figure~\eqref{fig:gibbs-burnin}
shows the sample energy and magnetization as a function of the number of burn-in
sweeps for the two Pegasus P6 instances used in the two dimensional experiments:
the random uniform instance (RAU) and the corrupted biased ferromagnet (CBFM).

For the RAU instance, the sample energy decreases during the first sweeps and then
approaches a stable plateau for all tested values of \(\beta_1\). The approach to the
plateau is faster at smaller \(\beta_1\), while the lowest temperature samples require
longer burn-in. The magnetization remains close to zero after burn-in, as expected for
a random instance without a global ferromagnetic bias. The absence of a systematic
late time drift in both observables indicates that the sampler has reached a stable
region for the low order observables used in the thermodynamic analysis.

For the CBFM instance, the energy also relaxes to a plateau after burn-in, but the
plateau values are lower and the magnetization remains finite. This finite
magnetization is expected because the CBFM contains a biased ferromagnetic
structure with local corruptions; it is not, by itself, a sign of poor sampling. As in the
RAU case, the slowest convergence occurs for the largest \(\beta_1\), where low energy
configurations are sampled more selectively. We therefore choose the production
burn-in from the plateau region of the energy and magnetization curves and use the
same criterion for all preparation temperatures.

These diagnostics do not prove that the full high dimensional distribution is exactly
Gibbsian. They show that the main observables controlling the initial energy scale
and the global bias are stationary after burn-in. In the main analysis, this check is
combined with pseudo-likelihood thermometry of the output samples and bootstrap
uncertainty estimates for the thermodynamic bounds. The resulting phase diagrams
should therefore be read as operational thermodynamic maps of the implemented
protocol, with accuracy limited by Gibbs sampling quality, hardware calibration, and
finite sampling statistics.

\section{D-Wave Advantage2 thermodynamic analysis}
\label{app:advantage2-results}
We report additional measurements obtained on a D-Wave Advantage2
processor for two dimensional RAU and CBFM instances.  The analysis follows the same protocol as in the main text. A classical initial state is
drawn from a Gibbs distribution of the programmed problem Hamiltonian at inverse
temperature \(\beta_1\). The system is then evolved under a reverse annealing cycle of
duration \(\tau=100~\mu{\rm s}\), with turning point \(s_p\), and the final bitstrings are
used to compute the two-point energy change
\[
\Delta E_1 =
E_p\!\left(\sigma(\tau)\right)
-
E_p\!\left(\sigma(0)\right).
\]
From the output configurations we estimate the effective inverse temperature
\(\beta_2\) by pseudo-likelihood thermometry and evaluate the same bounds on heat
and work used in the main analysis. The resulting temperature should be interpreted
as an effective temperature of the sampled Ising degrees of freedom. It is not a direct
measurement of the refrigerator plate temperature.
\begin{figure}[t!]
    \centering
    \includegraphics[width=\textwidth]{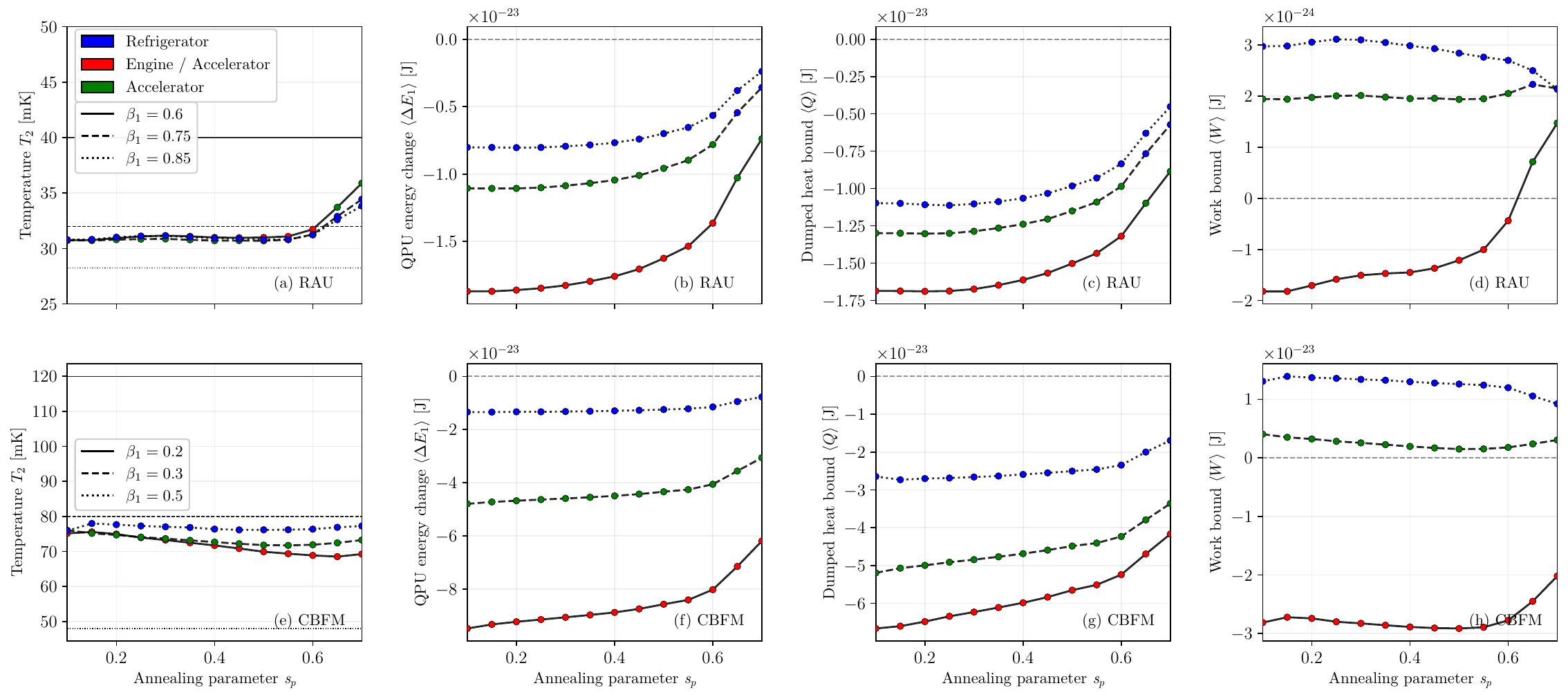}
    \caption{
    \textbf{D-Wave Advantage2 thermodynamic scans for two dimensional RAU and CBFM
    instances.}
    Panels (a--d) show the random uniform (RAU) instance for
    \(\beta_1=0.6,0.75,0.85\). Panels (e--h) show the corrupted biased ferromagnet
    (CBFM) for \(\beta_1=0.2,0.3,0.5\). The plotted quantities are the effective
    temperature \(T_2\), the mean processor energy change
    \(\langle \Delta E_1\rangle\), the bound on heat, and the bound on work as
    functions of the reverse annealing turning point \(s_p\), for a fixed cycle time
    \(\tau=100~\mu{\rm s}\). Point colours indicate the thermodynamic label assigned
    from the conservative sign classification used in the main text. The RAU instance
    gives effective temperatures near \(30\)--\(36~{\rm mK}\) over most of the scan,
    while the CBFM instance gives higher effective temperatures, approximately
    \(70\)--\(80~{\rm mK}\). The larger energy-change and heat-bound magnitudes in
    the CBFM case indicate a stronger dependence of the thermodynamic response on
    problem structure.
    }
    \label{fig:app-adv2-quantities}
\end{figure}
Figure~\ref{fig:app-adv2-quantities} summarizes the thermodynamic scans obtained
on Advantage2. For the RAU instance, the effective temperature \(T_2\) remains close
to \(31~{\rm mK}\) over most of the interval \(0.1\lesssim s_p\lesssim 0.55\), with a
moderate increase at larger turning points. In the same range, the mean processor
energy change \(\langle \Delta E_1\rangle\) is negative for the plotted initial
temperatures, and its magnitude decreases as \(s_p\) is increased. The heat bound
shows the same qualitative trend. The work bound is more sensitive to the initial
preparation: for the warmest RAU preparation shown here, \(\beta_1=0.6\), it is
negative over much of the scan and crosses toward positive values near the largest
turning points. For the colder preparations, \(\beta_1=0.75\) and \(0.85\), the work
bound is positive over most of the scan.

The CBFM instance shows a different thermodynamic response. Its inferred
effective temperature is higher, approximately \(70\)--\(80~{\rm mK}\), and varies only
weakly with \(s_p\). The magnitudes of
\(\langle \Delta E_1\rangle\) and of the heat bound are larger than in the RAU case.
This difference should not be interpreted as a direct change of the cryogenic bath
temperature. Rather, it indicates that the output ensemble of the CBFM instance is
described by a higher effective temperature under the programmed Hamiltonian and
the chosen reverse annealing protocol. This is consistent with the more structured
energy landscape of the CBFM problem, where biased domains and local corruptions
can make relaxation during the return branch less uniform than in a random uniform
instance.
\begin{figure}[t!]
    \centering
    \includegraphics[width=\textwidth]{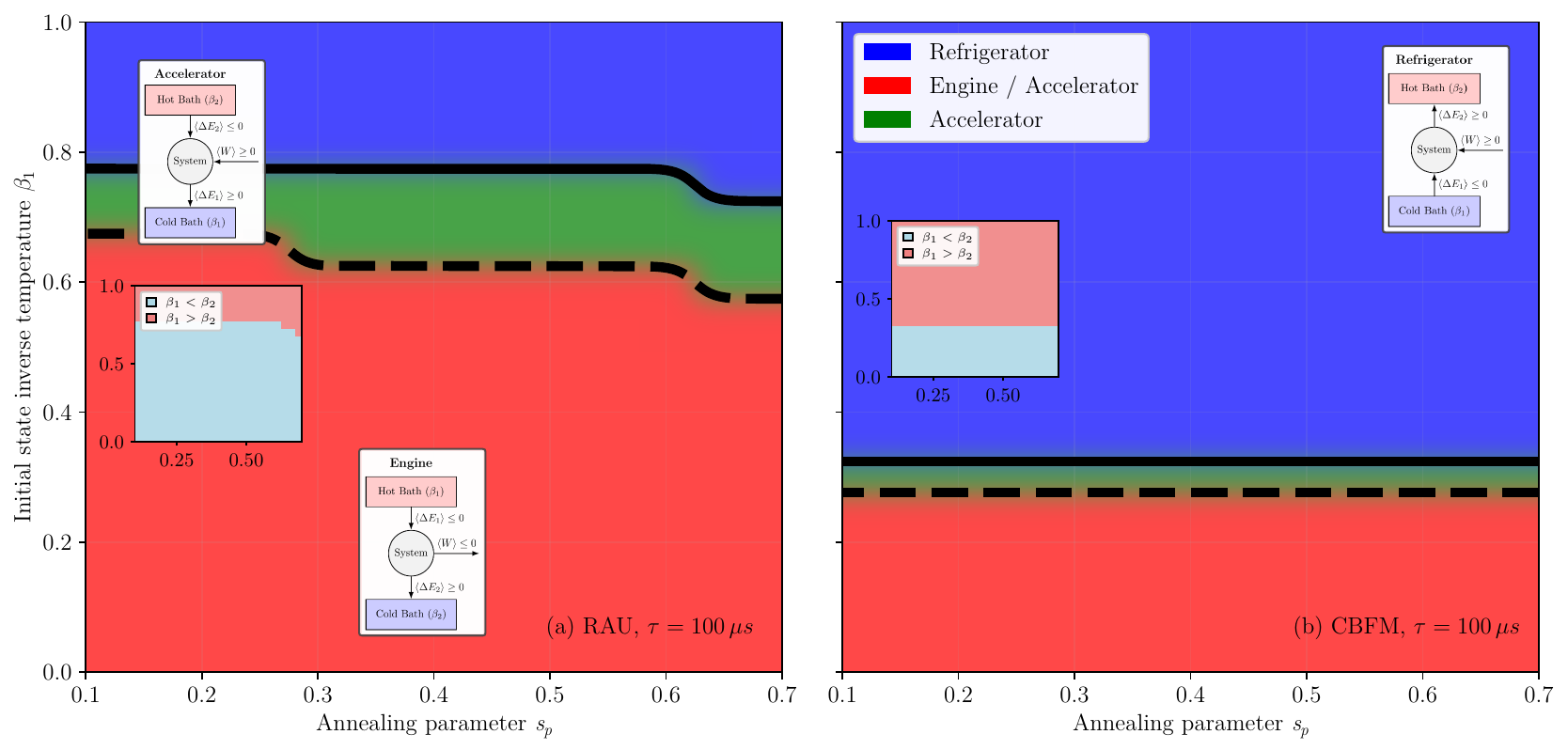}
    \caption{
    \textbf{Advantage2 operating mode maps for two dimensional RAU and CBFM
    instances.}
    The thermodynamic label is shown in the \((\beta_1,s_p)\) plane for
    \(\tau=100~\mu{\rm s}\). Panel (a) corresponds to the RAU instance and panel (b)
    to the CBFM instance. The colour code follows the conservative classification
    used in the main text: refrigerator, accelerator, and engine/accelerator-compatible
    regions. The dashed line indicates the condition \(\beta_1=\beta_2\) inferred from
    pseudo-likelihood thermometry. The RAU map contains a broad
    engine/accelerator-compatible region at lower \(\beta_1\), an intermediate
    accelerator band, and a refrigerator region at larger \(\beta_1\). The CBFM map
    shows the same ordering of regimes, but the boundaries are shifted to lower
    \(\beta_1\) and are nearly horizontal over most of the scanned \(s_p\) range.
    }
    \label{fig:app-adv2-phase}
\end{figure}
The corresponding phase diagrams are shown in
Figure~\eqref{fig:app-adv2-phase}. For the RAU instance, the operating modes are arranged
mainly by the initial inverse temperature. At lower \(\beta_1\), the data fall in the
engine/accelerator-compatible region. At intermediate \(\beta_1\), the system enters
an accelerator region. At larger \(\beta_1\), the classification becomes refrigerator.
The boundaries are approximately horizontal for shallow and intermediate turning
points, but bend downward when \(s_p\) approaches the upper end of the scanned
range. This indicates that the schedule depth affects the mode classification most
noticeably when the reverse annealing cycle remains closer to the final classical
Hamiltonian during the excursion.

For the CBFM instance, the same sequence of regimes is present, but the boundaries
are shifted to lower values of \(\beta_1\). The refrigerator region dominates for
\(\beta_1\) above roughly \(0.3\), while the engine/accelerator-compatible region lies
below this range. The accelerator band is comparatively narrow and remains close to
the boundary between these two regions. The weak dependence of the CBFM
boundaries on \(s_p\) suggests that, for this instance and cycle time, the initial
preparation temperature controls the thermodynamic label more strongly than the
precise turning point of the reverse annealing schedule.

The difference between RAU and CBFM is therefore useful for interpreting the
effective thermometry. The RAU instance gives a lower and nearly constant effective
temperature over most of the scan, whereas the CBFM instance gives a higher
effective temperature and larger energy-change bounds. These trends show that
\(T_2\) is an operational property of the full sampling protocol: it depends on the
programmed Hamiltonian, the embedding, the schedule, the hardware calibration, and
the finite time dynamics. It should not be read as a universal temperature of the
processor. The data support two conclusions. First, the
thermodynamic classification procedure transfers from Advantage to the Advantage2 processor without
requiring a change in the analysis. Second, the location of the operating mode
boundaries is problem dependent. The same hardware and the same cycle time can
give different effective temperatures and different heat and work bounds when the
programmed energy landscape is changed.

\section*{Data availability}
All study data are included in this article and the Supplementary Materials.
The datasets for the problems generated and analysed during the current study
are available from the following publicly accessible repository \href{https://github.com/iitis/pegasus-thermodynamic}{https://github.com/iitis/pegasus-thermodynamic}
\section*{Code availability}
The code is available from the following publicly accessible repository
\href{https://github.com/iitis/pegasus-thermodynamic}{https://github.com/iitis/pegasus-thermodynamic}

\bibliography{sn-bibliography}% common bib file

\section*{Acknowledgments}
The authors acknowledge the J\"ulich Supercomputing Centre for providing computing time on the D-Wave Advantage™ System JUPSI through the J\"ulich UNified Infrastructure for Quantum computing (JUNIQ). Z.M. acknowledges the hospitality of Forschungszentrum Jülich GmbH, funding from The Helmholtz Association through the Helmholtz Visiting Researcher Grant and the Ministry of Economic Affairs, Labour and Tourism Baden-Württemberg in the frame of the Competence Center Quantum Computing Baden-Württemberg (project ``KQCBW25''). B.G. and T.S. acknowledge Sonata Bis 10 project, No. 2020/38/E/ST3/00269. S.D. acknowledges support from the John Templeton Foundation under Grant No. 63626. Quantumz.io Sp. z o.o acknowledges support received from Polish Agency for Enterprise Development (PARP), Poland under Project No. FENG.01.01-IP.02-0625/23, titled Dynamic allocation of resources in industrial ecosystems susceptible to disturbances using physics-inspired algorithms and machine learning.
 
\section*{Author contributions}
Z.M. conceived the study, designed the research, and led the preparation of the manuscript. J.P. and T.Ś. developed the code, performed the experiments, analysed the data, and contributed to writing. F.J., B.G., and S.D. provided technical and scientific guidance, discussions, and supervision. All authors discussed the results, reviewed the manuscript, and approved the final version.
\section*{Competing interests}
The authors declare no competing interests.
\end{document}